\documentclass[twocolumn]{aastex63}

\usepackage{natbib}

\usepackage{graphicx}
\usepackage{epstopdf}
\usepackage{amsmath}
\usepackage{amssymb}
\usepackage{bm}
\usepackage{color}
\usepackage{multirow}
\usepackage{rotating}
\usepackage{makecell}

\def\gs{\mathrel{\raise0.35ex\hbox{$\scriptstyle >$}\kern-0.6em
\lower0.40ex\hbox{{$\scriptstyle \sim$}}}}
\def\ls{\mathrel{\raise0.35ex\hbox{$\scriptstyle <$}\kern-0.6em
\lower0.40ex\hbox{{$\scriptstyle \sim$}}}}

\newcommand{\Hb}{H$\beta$}
\newcommand{\Ha}{H$\alpha$}

\newcommand{\heii}{He\,{\sc ii}}

\newcommand{\civ}{C\,{\sc iv}}
\newcommand{\civdub}{C\,{\sc iv} $\lambda\lambda$1548,1550}
\newcommand{\ciii}{C\,{\sc iii}]}

\newcommand{\oiiisemi}{O\,{\sc iii}]}

\newcommand{\mgii}{Mg\,{\sc ii}}
\newcommand{\siivoiv}{Si\,{\sc iv}+O\,{\sc iv}]}
\newcommand{\niv}{N\,{\sc iv}]}
\newcommand{\niii}{N\,{\sc iii}]}

\shorttitle{Reverberation mapping of SDSS J2222+2745} 
\shortauthors{Williams et al.}
 
\begin{document}

\title{The black hole mass of the $z=2.805$ multiply imaged quasar SDSS J2222+2745 from velocity-resolved time lags of the \civ\ emission line}

\author[0000-0002-4645-6578]{Peter R. Williams}
\affiliation{Department of Physics and Astronomy, University of California, Los Angeles, CA 90095-1547, USA}

\author[0000-0002-8460-0390]{Tommaso Treu}
\altaffiliation{Packard Fellow}
\affiliation{Department of Physics and Astronomy, University of California, Los Angeles, CA 90095-1547, USA}

\author[0000-0003-2200-5606]{H\r{a}kon Dahle}
\affiliation{Institute of Theoretical Astrophysics, University of Oslo, PO Box 1029, Blindern 0315, Oslo, Norway}

\author[0000-0001-8818-0795]{Stefano Valenti}
\affiliation{Department of Physics, University of California, Davis, CA 95616, USA}

\author[0000-0002-8860-1032]{Louis Abramson}
\affiliation{The Observatories of the Carnegie Institution for Science, 813 Santa Barbara St., Pasadena, CA 91101, USA}

\author[0000-0002-3026-0562]{Aaron J. Barth}
\affiliation{Department of Physics and Astronomy, University of California at Irvine, 4129 Frederick Reines Hall, Irvine, CA 92697-4575, USA}

\author[0000-0003-1370-5010]{Michael Gladders}
\affiliation{Department of Astronomy \& Astrophysics, The University of Chicago, 5640 S. Ellis Avenue, Chicago, IL 60637, USA}

\author[0000-0003-1728-0304]{Keith Horne}
\affiliation{SUPA Physics and Astronomy, University of St. Andrews, Fife, KY16 9SS, UK}

\author[0000-0002-7559-0864]{Keren Sharon}
\affiliation{Department of Astronomy, University of Michigan, 1085 S. University Avenue, Ann Arbor, MI 48109, USA}

\correspondingauthor{Peter R. Williams}
\email{pwilliams@astro.ucla.edu}

\begin{abstract}
  We present the first results of a 4.5 year monitoring campaign of the three bright images of multiply imaged $z=2.805$ quasar SDSS J2222+2745 using the Gemini North Multi-Object Spectrograph (GMOS-N) and the Nordic Optical Telescope (NOT).
  We take advantage of gravitational time delays to construct light curves surpassing 6 years in duration and achieve average spectroscopic cadence of 10 days during the 8 months of visibility per season.
  Using multiple secondary calibrators and advanced reduction techniques, we achieve percent-level spectrophotometric precision and
  carry out an unprecedented reverberation mapping analysis, measuring both integrated and velocity-resolved time lags for \civ.
  The full line lags the continuum by $\tau_{\rm cen} = 36.5^{+2.9}_{-3.9}$ rest-frame days.
  We combine our measurement with published \civ\ lags and derive the $r_{\rm BLR}-L$ relationship $\log_{10}( \tau / {\rm day}) = (1.00\pm0.08) + (0.48\pm0.04) \log_{10} [\lambda L_\lambda(1350{\rm \AA})/10^{44}~{\rm erg~s}^{-1}]$ with 0.32$\pm$0.06 dex intrinsic scatter.
  The velocity-resolved lags are consistent with circular Keplerian orbits,
  with $\tau_{\rm cen} = 86.2^{+4.5}_{-5.0}$, $25^{+11}_{-15}$, and $7.5^{+4.2}_{-3.5}$ rest-frame days for the core, blue wing, and red wing, respectively.
  Using $\sigma_{\rm line}$ with the mean spectrum and assuming $\log_{10} (f_{{\rm mean},\sigma}) = 0.52 \pm 0.26$, we derive $\log_{10}(M_{\rm BH}/M_{\odot}) = 8.63 \pm 0.27$.
  Given the quality of the data, this system represents a unique benchmark for calibration of $M_{\rm BH}$ estimators at high redshift.
  Future work will present dynamical modeling of the data to constrain the virial factor $f$ and $M_{\rm BH}$. 
\end{abstract}



\section{Introduction}
\label{sect:intro}
The ability to measure precise black hole masses is critical to understanding the formation and accretion history of supermassive black holes (SMBHs) and their role in the evolution of their host galaxies over cosmic time.
While stellar and gas kinematics can be used in nearby galaxies \citep[e.g.,][]{kormendy95, ferrarese05, McConnell+13}, these approaches are not feasible in the distant universe where the black hole sphere of influence cannot be resolved, with the exception of extraordinary cases \citep[e.g., ][]{Gravity++18}.
The technique of reverberation mapping \citep{blandford82, peterson93, peterson14} utilizes the gaseous broad emission line region (BLR) in the inner light days of active galactic nuclei (AGNs) to make $M_{\rm BH}$ measurements possible at cosmological distances.

Given its proximity to the central black hole, the BLR gas moves at speeds on the order of $10{,}000~{\rm km~s}^{-1}$, leading to a Doppler broadening of emission lines that can be measured, $\Delta V$.
By monitoring the AGN over time, one can measure a time lag, $\tau$, between fluctuations in the AGN continuum and the broad emission line strength.
Assuming the motion of the gas is dominated by the black hole's gravity, these quantities provide an estimate of the black hole mass,
\begin{align}
\label{eqn:rm}
M_{\rm BH} = f\frac{c\tau \Delta V^2}{G},
\end{align}
where $f$ is a dimensionless virial factor of order unity that accounts for the unknown structure, kinematics, and orientation of the BLR.

The observed relation between the BLR radius and AGN luminosity \citep{kaspi00,Kaspi++05,bentz13} has enabled ``single-epoch'' $M_{\rm BH}$ measurements in which $\tau$ and $\Delta V$ can be measured from a single spectrum.
This opens up the possibility of using large spectroscopic surveys to measure thousands of black hole masses and study the co-evolution of black holes and their host galaxies across cosmic time \citep[e.g.,][]{Shen++11}.
However, the majority of reverberation mapping measurements are based on the \Hb\ emission line which is at optical wavelengths for local AGNs, making it suitable for ground-based campaigns.
To extend the single-epoch method to higher redshifts requires the use of UV emission lines---most commonly \civ\ \citep{Vestergaard02, Vest+06} and \mgii\ \citep{McLure+02}.
However, the relationship between the \Hb, \civ, and \mgii\ BLRs is poorly understood, and the limited redshift and luminosity range of UV reverberation mapping measurements means that the $r_{\rm BLR}-L$ relationships for these lines rely on extrapolations from more local AGNs.
Additionally, differences between \civ\ and \Hb\ emission line profiles have called into question the validity of \civ\ as a single-epoch mass estimator \citep[e.g.,][]{Baskin+05, Shen+12}, although other analyses have suggested that these issues are mitigated by proper data quality selection \citep[e.g.,][]{Vest+06}.

A number of campaigns have aimed to improve the \civ\ $r_{\rm BLR}-L$ relation by increasing the size and dynamic range of the sample of UV-based reverberation mapping $M_{\rm BH}$ measurements \citep{Peterson++05, Peterson++06, Kaspi++07, Trevese++14, Lira++18, Lira++20, Hoormann++19, Grier++19}, but such measurements are complicated for a number of reasons.
First, high-$z$ measurements typically focus on high luminosity AGNs to reach the required signal-to-noise ratio of the spectra.
These AGNs have large BLRs due to the $r_{\rm BLR} - L$ relation, requiring long campaign durations, an obstacle that is only amplified by the cosmological $(1+z)$ time dilation.
Additionally, high-luminosity AGNs tend to have smaller-amplitude fluctuations in the continuum compared to their lower-luminosity counterparts, but large fluctuations with distinct features are essential for measuring emission line time lags.

In this paper, we present an extraordinary object that solves all of these issues by means of strong gravitational lensing.
SDSS J2222+2745 is a quasar at redshift $z=2.805$ that is lensed by a foreground galaxy cluster, discovered by \citet{Dahle++13} as part of the Sloan Giant Arcs Survey \citep{Hennawi++08}\footnote{The name SDSS J2222+2745 is used in some publications to refer to the foreground galaxy cluster or the full lens system. The quasar can be found on the NASA/IPAC Extragalactic Database (NED) with the identifier SDSS J2222+2745:[SBD2017]}. 
Due to the image magnifications \cite[$\mu_A=14.5\pm 2.7$, $\mu_B=10.8\pm 4.3$, $\mu_C=6.7\pm 1.0$;][]{Sharon++17}, the quasar is visible at $g\sim21$ despite being intrinsically dimmer, and the estimated \civ\ time lag is on the order of 100 days in the observed frame.
Photometric monitoring of the brightest images A, B, and C has shown that image C leads the others by two years \citep{Dahle++15}, meaning all light curves can be extended by two years, shortening the required campaign duration further.
Finally, the leading image C was observed to undergo extreme flux variations between 2014 and 2016, brightening by over one magnitude in the $g$ band, classifying it as an ``extreme variability quasar'' \citep[EVQ,][]{Rumbaugh++18}.
EVQs are rarely found at $z\sim 3$ in part due to the high luminosities required for them to be observable from Earth, but the magnification effect of gravitational lensing allows us to overcome this obstacle.
Since the trailing images necessarily follow the behavior of the leading image, they were guaranteed to undergo the same large-scale flux variations that are necessary for reverberation mapping measurements.

In Sections~\ref{sect:observations} and~3, we describe the spectroscopic and photometric monitoring campaigns and the procedures for data reduction and flux calibration. 
In Section 4, we describe the multicomponent model used to decompose the spectra into their individual components, and in Section 5 we describe the intercalibration between the spectroscopic and photometric emission line measurements.
In Section 6, we present spectroscopic and photometric \civ\ and continuum light curves and measure velocity-resolved time lags, placing SDSS J2222+2745 on the \civ\ $r_{\rm BLR}-L$ relation.
We conclude in Section 7.
When necessary, we adopt a $\Lambda$CDM cosmology with $H_0 = 70~{\rm km~s}^{-1}~{\rm Mpc}^{-1}$, $\Omega_M = 0.3$, and $\Omega_\Lambda = 0.7$.


\section{Observations}
\label{sect:observations}

Spectroscopic observations were carried out using the Multi-Object Spectrograph at Gemini Observatory North \citep[GMOS-N;][]{GMOS}.
The spectra were taken in queue mode, with 6000 seconds of exposure time per lunation, between April and December when the target was visible.
From June 2016 - December 2016, observations consisted of four 1500 second exposures.
In 2017, GMOS-N upgraded the CCDs from the previous e2v deep depletion (DD) device detectors to the current Hamamatsu detector array.
The Hamamatsu detectors are physically thicker than the e2v DD detectors, and therefore have a higher cosmic ray hit rate, so we changed the observing strategy to use six 1000 second exposures for the duration of the campaign.
The minimum observing conditions provided to the queue system were 80th percentile sky background (background $V$-band magnitude $\mu_V > 19.5$), 70th percentile cloud cover (${\rm signal~loss}<20\%$), and 70th percentile image quality (FWHM of image at $0.475~\mu{\rm m} < 0.35^{\prime\prime}$)\footnote{A detailed explanation of the observing conditions can be found at \url{https://www.gemini.edu/observing/telescopes-and-sites/sites\#Constraints}}.
All timing windows were scheduled at least three days from the full moon, and shorter windows were set when the target was close to the moon.
The observing constraints were loosened and observations were allowed to be split across multiple nights during extended periods of unfavorable conditions.
This was particularly important during the Fall 2017 months when weather on Maunakea was particularly bad, as well as Spring 2020 when the telescopes were operating at limited capacity due to the COVID-19 pandemic.

We designed the slit mask with slitlets on the brightest quasar images A, B, and C, and placed additional slitlets on three bright nearby stars to be used for flux calibration.
The slitlets were 2 arcseconds wide, which is much wider than the expected seeing and residual mis-alignments, minimizing slit losses.
We used the B600 grating, with 600 lines mm$^{-1}$, centered at 6500\AA, and the GG455 order blocking filter.
After dithering and flux calibration, this setup gives coverage from $\sim$5000 to 8200 \AA\ at 1.0 \AA\ pixel$^{-1}$ with $2\times2$ binning (0.9 \AA\ pixel$^{-1}$ for the e2v DD detectors.)
These choices provide simultaneous coverage of both the \civdub\ and \ciii\ $\lambda$1909 emission lines.
Calibration frames were taken every night that data were taken, including bias, flat, and CuAr arc lamp exposures.

We missed one full lunation in August 2017 and 1/3 of our time allocation in July 2019 due to poor weather.
We also missed the April 2020 lunation due to the COVID-19 telescope shutdown, but were able to use the time to get an additional epoch in July 2020.
In total, we obtained spectra for 36 lunations from June 2016 to September 2020.
Nine of these were split across two nights and one was split across three nights, giving a total of 47 epochs.
The median signal-to-noise ($S/N$) ratios per pixel of the spectra after flux calibration are 30.7, 28.4, and 23.6 for images A, B, and C, respectively.
The standard deviations of the distributions of $S/N$ per pixel over the duration of the campaign were 6.0, 5.7, and 7.1.

In addition to spectroscopy, we obtained higher cadence photometric observations with the Alhambra Faint Object Spectrograph and Camera (ALFOSC) at the 2.56m Nordic Optical Telescope (NOT).
Observations were taken with the SDSS\_g$^\prime$, SDSS\_i$^\prime$, and HeI 588\_6 (hereafter the $g$-, $i$-, and narrow-band) filters with mean cadences of 18.3, 18.7, and 17.7 days for the $g$, $i$, and narrow band, respectively.
The narrow-band filter covers the core of the redshifted \civ\ emission line, providing higher-cadence measurements of the emission line flux to supplement the spectroscopy.
Each night consisted of three 600 second exposures with the $g$-band and narrow-band filters and three 300 second for the $i$-band filter.

The $g$-band photometric monitoring began in September 2011, while the narrow and $i$-band monitoring did not begin until June and July 2016, respectively.
The $g$-, $i$-, and narrow-band photometry used in this paper extend until September 2020, December 2019, and October 2019, respectively. 
Beginning 2016 June 1, we obtained 86, 49, and 41 epochs for the $g$, $i$, and narrow band.
The median $S/N$ $\pm$ standard deviation in $S/N$ over the duration of the campaign were $(g, i, {\rm narrow}) = (52 \pm 24, 33.7\pm 9.7, 24.5\pm 8.0)$ for image A,  $(42\pm 19, 26.6\pm 7.6, 18.7\pm 6.8)$ for image B, and $(40\pm 17, 23.7\pm 7.9, 18.6\pm 7.0)$ for image C.


\section{Spectral data processing}
\label{sect:reduction}

Data processing began with bias subtraction, flat-fielding, and wavelength calibration using the \texttt{GMOS} {\sc IRAF\footnote{IRAF was distributed by the National Optical Astronomy Observatory, which was managed by the Association of Universities for Research in Astronomy (AURA) under a cooperative agreement with the National Science Foundation.}} routines provided by Gemini Observatory.
An additional quantum efficiency (QE) correction step was added for all data taken after 2016, due to the different QEs of the Hamamatsu detector chips.
Following these steps, we use a {\sc python} implementation of L.A.Cosmic \citep{lacosmic} to perform cosmic ray cleaning on the 2D spectra.
The quasar and standard star spectra are then extracted using the {\sc IRAF} \texttt{apall} routine, using an unweighted extraction with a fixed 2.74$^{\prime\prime}$ aperture.

\subsection{Flux calibration}
For each epoch we obtain observations of three standard stars for the purpose of flux calibration.
The stars are in three separate slits of the slitmask and the observations are obtained simultaneously with the quasar observations.
These stars have not been previously studied, so we use a spectral type fitter \citep{Pickles+10}\footnote{\url{https://lco.global/~apickles/SpecMatch/}} with the Sloan Digital Sky Survey \citep[SDSS,][]{SDSS} photometry to determine their spectral types.
We then obtain a template spectrum based on that spectral type from the The Indo-U.S. Library of Coud\'e Feed Stellar Spectra \citep{Valdes++04}\footnote{\url{https://www.noao.edu/cflib/}}.
The SDSS photometry used and the spectral type determined with the fitter are listed in Table \ref{table:standard_properties}.

\begin{deluxetable*}{lccccccccc}
\tabletypesize{\small}
\tablecaption{Standard star properties}
\tablewidth{0pt}
\tablehead{ 
\colhead{Star} & 
\colhead{R.A.} &
\colhead{Dec.} &
\colhead{$u$} & 
\colhead{$g$} & 
\colhead{$r$} & 
\colhead{$i$} & 
\colhead{$z$} & 
\colhead{Spectral}  &
\colhead{Template} \\[-5pt]
\colhead{} &
\colhead{(J2000.0)} &
\colhead{(J2000.0)} &
\colhead{(mag)} &
\colhead{(mag)} &
\colhead{(mag)} &
\colhead{(mag)} &
\colhead{(mag)} &
\colhead{type} &
\colhead{file}
}
\startdata
stand1	& 22:22:04.30 & +27:45:34.0 & $ 20.80 \pm 0.06 $	&	$ 19.66 \pm 0.01 $	&	$ 19.03 \pm 0.01 $	&	$ 18.74 \pm 0.01 $	&	$ 18.59 \pm 0.03 $	&	K0V	&	G 149-25	\\
stand2	& 22:22:06.23 & +27:45:35.6 & $ 20.22 \pm 0.04 $	&	$ 19.04 \pm 0.01 $	&	$ 18.56 \pm 0.01 $	&	$ 18.35 \pm 0.01 $	&	$ 18.27 \pm 0.02 $	&	G5V	&	G 11-45	\\
stand3	& 22:22:05.36 & +27:45:37.9 & 	$ 21.67 \pm 0.12 $	&	$ 19.78 \pm 0.01 $	&	$ 18.96 \pm 0.01 $	&	$ 18.67 \pm 0.01 $	&	$ 18.49 \pm 0.03 $	&	K3IV	&	121146	
\enddata
\tablecomments{The three standard stars used for flux calibration along with their coordinates, SDSS photometry, fitted spectral type, and the Indo-U.S. Library of Coud\'e Feed Stellar Spectra template file names. 
\label{table:standard_properties}}
\end{deluxetable*} 

Using the normalized template spectra and the SDSS magnitudes, we can compute a response function, $r(\lambda)$, to convert the observed spectra from units of ${\rm counts~sec}^{-1}$ to physical units of ${\rm erg~s}^{-1}~{\rm cm}^{-2}~{\rm \AA}^{-1}$.
The response function accounts for the CCD sensitivity as well as the transparency of the atmosphere as a function of wavelength.
Assuming we know the true spectrum of the standard stars in flux units, this is simply
\begin{align}
r(\lambda) = {\rm template}(\lambda) / {\rm star}(\lambda).
\end{align}
Prior to computing $r(\lambda)$, we need to fine tune the wavelength solution, correct for the radial velocities of the standard stars, and convolve the template spectra by the point spread function (PSF) of the observed spectra.
These steps are described below.

\subsubsection{Slit positioning correction}
\label{sect:slitposition}
Since the slit width is much larger than the PSF, the positions of the quasar images and stars on the slit can affect the position in the wavelength dispersion direction on the CCD.
If the radial velocities of the standard stars were known, we could use the stellar absorption features to determine this wavelength offset, but these values have not yet been measured.

\begin{figure}[h!]
\begin{center}
\includegraphics[width=3.3in]{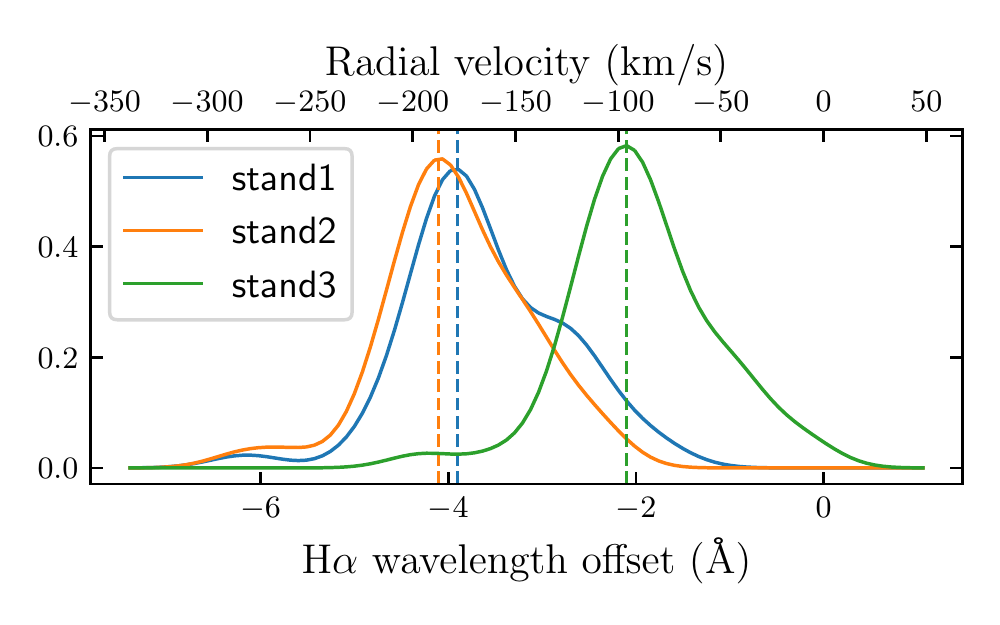}  
\caption{Distributions of the position of the \Ha\ absorption line, relative to 6563~\AA\ and the corresponding radial velocity for each of the three standard stars.
Note that these are not necessarily true radial velocities since they include a zero-point offset from the positioning of the stars on the slit.
\label{fig:rv_distribution}}
\end{center}
\end{figure}

Instead, we measure the position of the sharp \Ha\ absorption line for each standard star.
Figure \ref{fig:rv_distribution} shows the distribution of the measured wavelength offsets relative to $\lambda = 6563~{\rm \AA}$ and the corresponding radial velocity.
These are not necessarily the radial velocities of the stars relative to the heliocentric frame since the wavelength offset includes a zero-point offset due to the positioning on the slit.
Using the median values of these distributions, we assign for each star a ``true'' radial velocity for calibration purposes and determine the correct position of the \Ha\ line based on this value.
We then add a fixed value in angstroms to each wavelength in the spectra so that \Ha\ is aligned to this reference wavelength, correcting for any offsets in slit positioning.

\begin{figure}[h!]
\begin{center}
\includegraphics[width=3.3in]{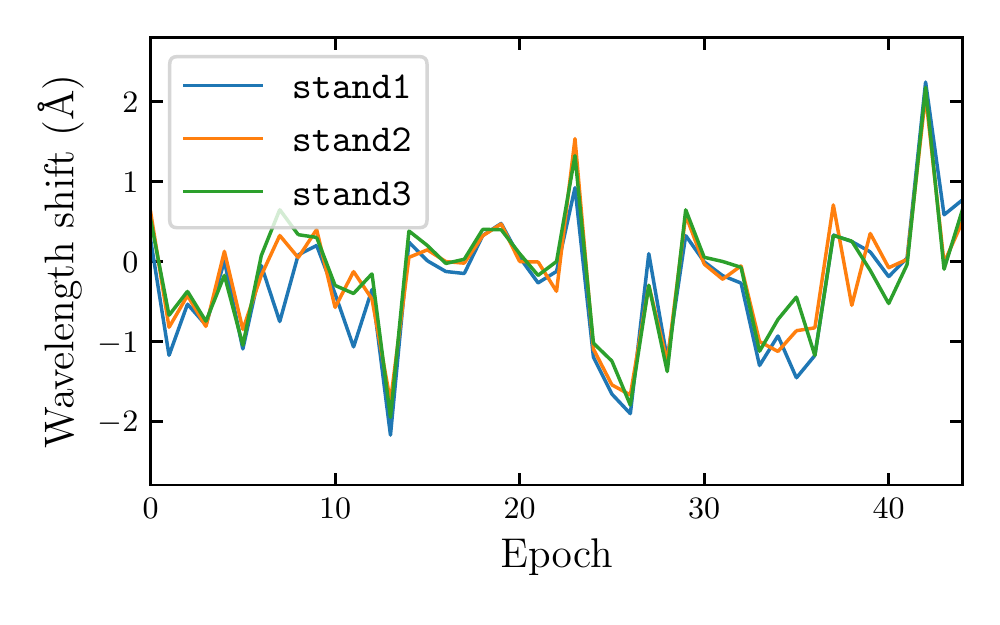}  
\caption{Wavelength offsets for each standard star to correct for variations in the positioning of the standard stars on the slits.
\label{fig:slit_positioning_offset}}
\end{center}
\end{figure}

Assuming the positioning deviations in the slit are due only to linear shifts of the slitmask and not due to rotations, the shift should be the same for every object on the CCD.
Figure \ref{fig:slit_positioning_offset} shows the wavelength shifts required for each of the three standard stars over the duration of the observing campaign.
The wavelength shifts at each epoch are very similar for all three stars, as expected.
We can therefore use these offsets to apply a shift to the quasar spectra, which should be affected in the same way.
We calculate the mean of the offsets for the three standard stars and shift the three quasar spectra by this amount.

\subsubsection{Radial velocity and PSF correction}
Before calculating the response function, we need to shift the template spectra according to the stellar radial velocities as well as convolve the templates with a point spread function (PSF) to match that of the observations.
We use the radial velocities determined in the previous section, and the PSF determination is done as follows:

For simplicity, we assume that the PSF is Gaussian in wavelength with standard deviation, $\sigma_{\rm PSF}$.
We also assume that $\sigma_{\rm PSF}$ is constant over the wavelength range of the spectra.
After multiplying the wavelength axis of the template spectrum by a factor $\gamma = 1 + v/c$ to correct for the radial velocity of the star, we multiply the template by a pseudo-response function, $\alpha + \beta \lambda$, and convolve with a Gaussian $\mathcal{N}(0,\sigma_{\rm PSF}^2)$.

\begin{figure}[h!]
\begin{center}
\includegraphics[width=3.3in]{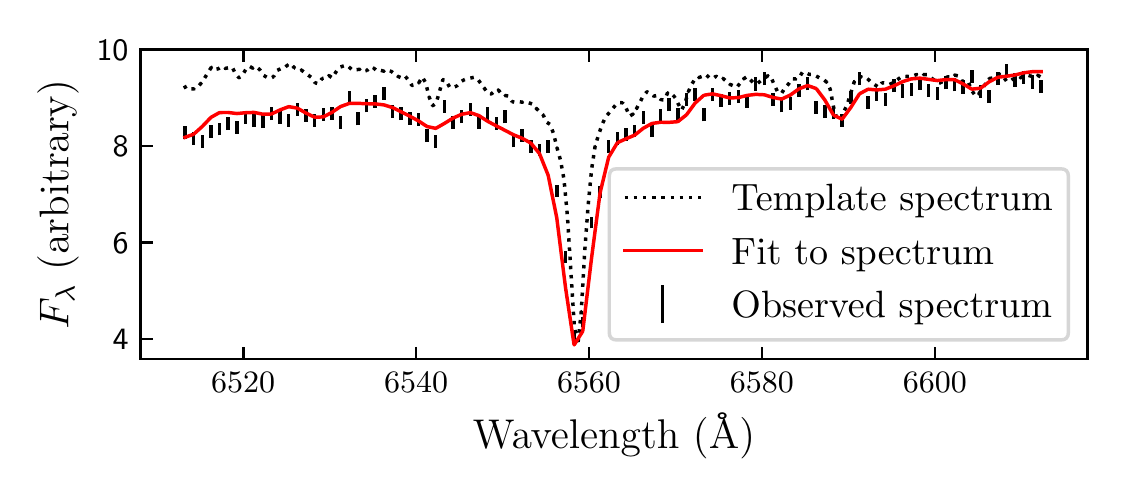}  
\caption{Standard star template spectrum corrected for radial velocity, scaled to match the observed flux values, and convolved to match the observed H$\alpha$ width.
Note that the offset in the \Ha\ trough from $\lambda = 6563~{\rm \AA}$ is due to the radial velocity of the observed star to which the template has been aligned.
\label{fig:template_model_fit}}
\end{center}
\end{figure}

We then use the minimizer \texttt{scipy.optimize.minimize} \citep{scipy} to find the parameters $\theta = (\alpha, \beta, \sigma_{\rm PSF})$ that maximize the log-likelihood
\begin{multline}
l = -\frac{1}{2} \sum_{\lambda \in [\lambda_{min},\lambda_{max}]} \Bigg[\frac{(f_{\rm obs}(\lambda) - M(\lambda, \theta))^2}{\sigma_{\rm obs}^2(\lambda)} + \\
\ln\left(2 \pi \sigma_{\rm obs}^2(\lambda)\right)\Bigg],
\end{multline}
where $f_{\rm obs}$ is the observed spectrum flux, $\sigma_{\rm obs}$ is the observed flux uncertainty, and $M$ is our model
\begin{align}
M(\lambda; \alpha, \beta, \sigma_{\rm PSF}^2) = \left[(\alpha + \beta \lambda)f_{\rm temp}(\lambda)\right] \ast \mathcal{N}(0,\sigma_{\rm PSF}^2), 
\end{align}
where $f_{\rm temp}$ is the radial-velocity-corrected template.
We use a 50 \AA~radius window centered on the sharp H$\alpha$ absorption feature to do the fitting.
An example fit is shown in Figure \ref{fig:template_model_fit}.

\subsubsection{Calculating the response function}
\label{sect:resp_calc}
To calculate the response function, the template spectra are first shifted in wavelength according to the radial velocities determined in the previous step, and then convolved with a Gaussian of width $\sigma_{\rm PSF}$.
The spectra are then linearly interpolated onto the wavelength scale of the observed standard star spectra.
The response is then simply calculated, pixel by pixel, as $r(\lambda_i) = f_{\rm temp}(\lambda_i)/f_{\rm stand}(\lambda_i)$.
Due to wavelength-dependent offsets in the three response functions arising from imperfect standard star template matches, we choose to use only the brightest standard star to compute the response function. We use the other two stars to perform sanity checks and estimate the uncertainty in the calibration.

\subsection{Uncertainty due to flux calibration}
\label{sect:calib_uncertainty}
The flux calibration procedure described above has the potential to introduce additional uncertainty that is not accounted for in the formal propagation of uncertainty due to random noise.
Photometric monitoring confirms that the standard stars do not vary significantly over the course of the observing campaign, so we can use them to estimate the magnitude of this additional calibration uncertainty.

\begin{figure}[h!]
\begin{center}
\includegraphics[width=3in]{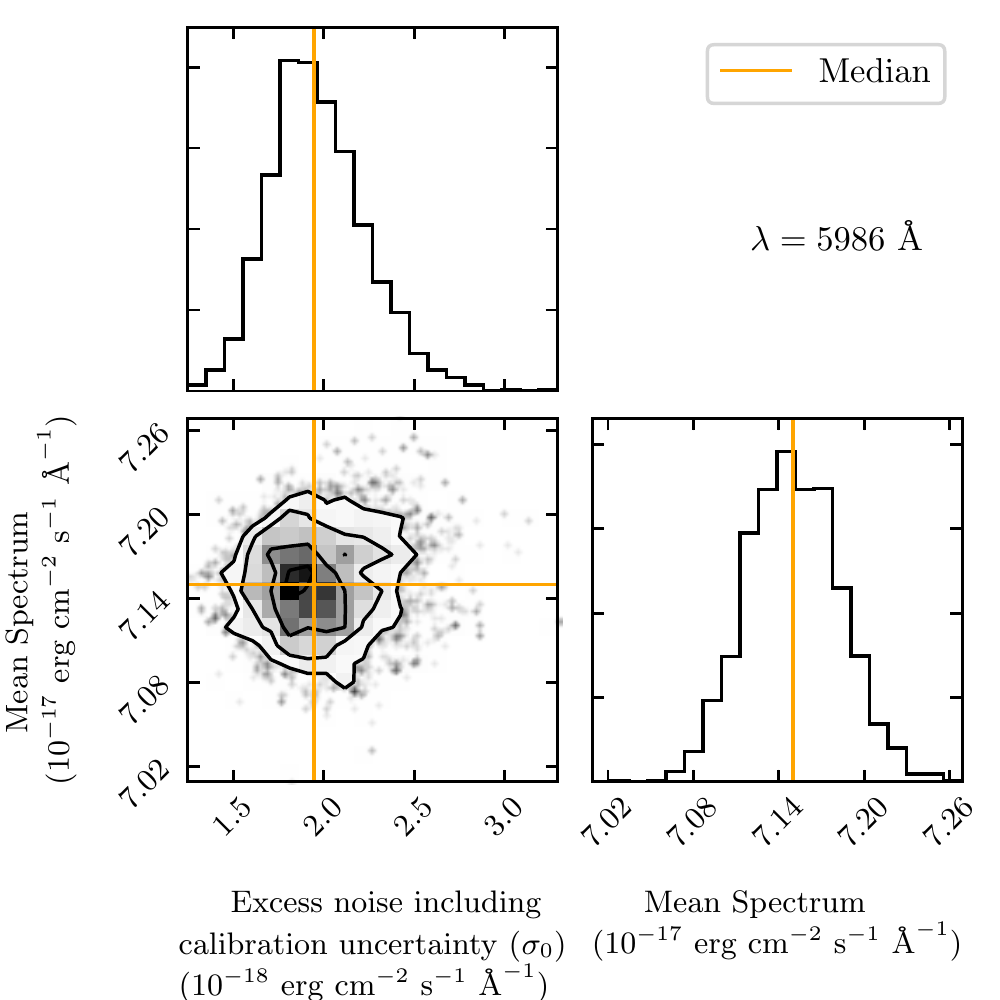}  
\caption{Corner plot showing the inference on the mean standard star 1 spectrum and the additional uncertainty due to calibration and other systematics. The example shown is for $\lambda=5986$ \AA, which falls on the \civ\ emission line. The median values are shown by the orange lines.
\label{fig:calib_uncertainty_corner}}
\end{center}
\end{figure}

\begin{figure*}[t!]
\begin{center}
\includegraphics[width=6in]{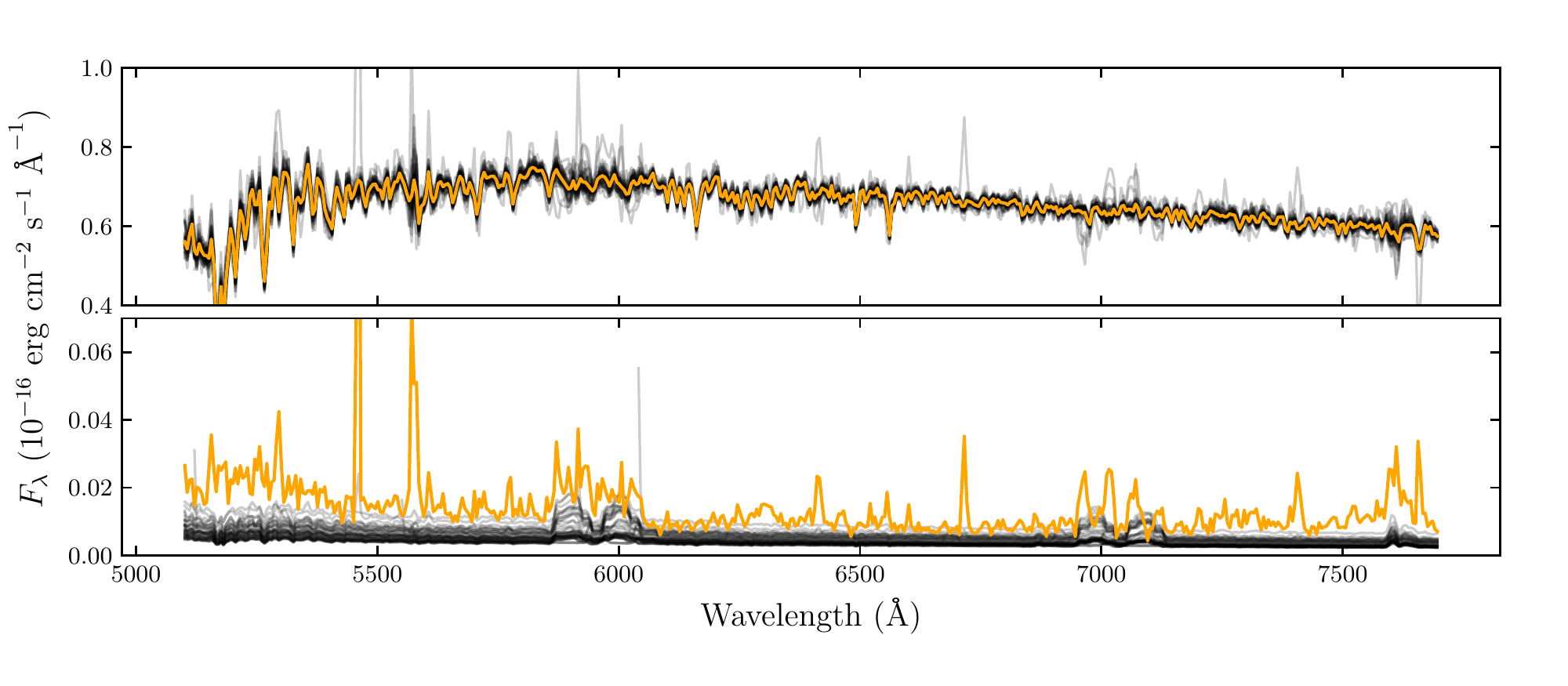}  
\caption{\textit{Upper panel}: Inferred mean spectrum $m(\lambda)$ (orange) and flux calibrated individual spectra (grey) for the standard star \texttt{stand1}. \textit{Bottom panel}: Inferred $\sigma_0(\lambda)$ (orange) and the individual spectra's random uncertainty (grey). The spectra shown here have been re-binned to wavelength bins of 5 \AA.
\label{fig:calib_uncertainty_spec}}
\end{center}
\end{figure*}

We assume that the standard star spectrum remains constant over time with some mean spectrum, $m(\lambda)$.
The uncertainty in the spectra can be described as a combination of the random noise for each epoch, $\sigma_i(\lambda)$, and an additional uncertainty source due to flux calibration, $\sigma_0(\lambda)$. Combined, the total uncertainty for each epoch is $\sigma_{{\rm tot},i}^2(\lambda) = \sigma_i^2(\lambda) + \sigma_0^2(\lambda)$.

The response function is calculated with the brightest of the standard stars, \texttt{stand2}, so we use the other stars, \texttt{stand1} and \texttt{stand3}, with the Markov chain Monte-Carlo (MCMC) code {\sc emcee} \citep{Foreman-Mackey++13} to explore the parameter space of $\sigma_0(\lambda)$ and $m(\lambda)$ and quantify the additional uncertainty due to the flux calibration.
The log-likelihood function is
\begin{align}
l(\lambda) = -\frac{1}{2} \sum_{i} \Bigg[\frac{(d_i(\lambda) - m(\lambda))^2}{\sigma_{{\rm tot},i}^2(\lambda)} + \ln\left(2 \pi \sigma_{{\rm tot},i}^2(\lambda)\right)\Bigg],
\end{align}
where $d_i(\lambda)$ is the observed flux calibrated standard star spectrum and the sum is over all epochs.
From the resulting MCMC samples, we can estimate the value of $\sigma_0(\lambda)$ for each wavelength bin.

Using the full wavelength resolution, $\sigma_0$ is too small compared to the random uncertainty to be measurable.
In order to make a measurement, we increase the $S/N$ by binning the spectra into wavelength bins of 5 \AA~width.
A corner plot showing the posterior distribution for one of the wavelength bins is shown in Figure \ref{fig:calib_uncertainty_corner}.
The orange lines show the median values for the two parameters $\sigma_0(\lambda)$ and $m(\lambda)$.

Figure \ref{fig:calib_uncertainty_spec} shows $m(\lambda)$ and $\sigma_0(\lambda)$ along with the binned spectra and $\sigma_i$ for each of the individual epochs.
Note that the spectra have been binned by a factor of 5 in the wavelength scale, so the random uncertainties are reduced by a factor of $1/\sqrt{5}$.
Thus, on the un-binned wavelength scale, the random uncertainty still dominates over the calibration uncertainty for most wavelengths.

\begin{figure*}[t!]
\begin{center}
\includegraphics[width=6in]{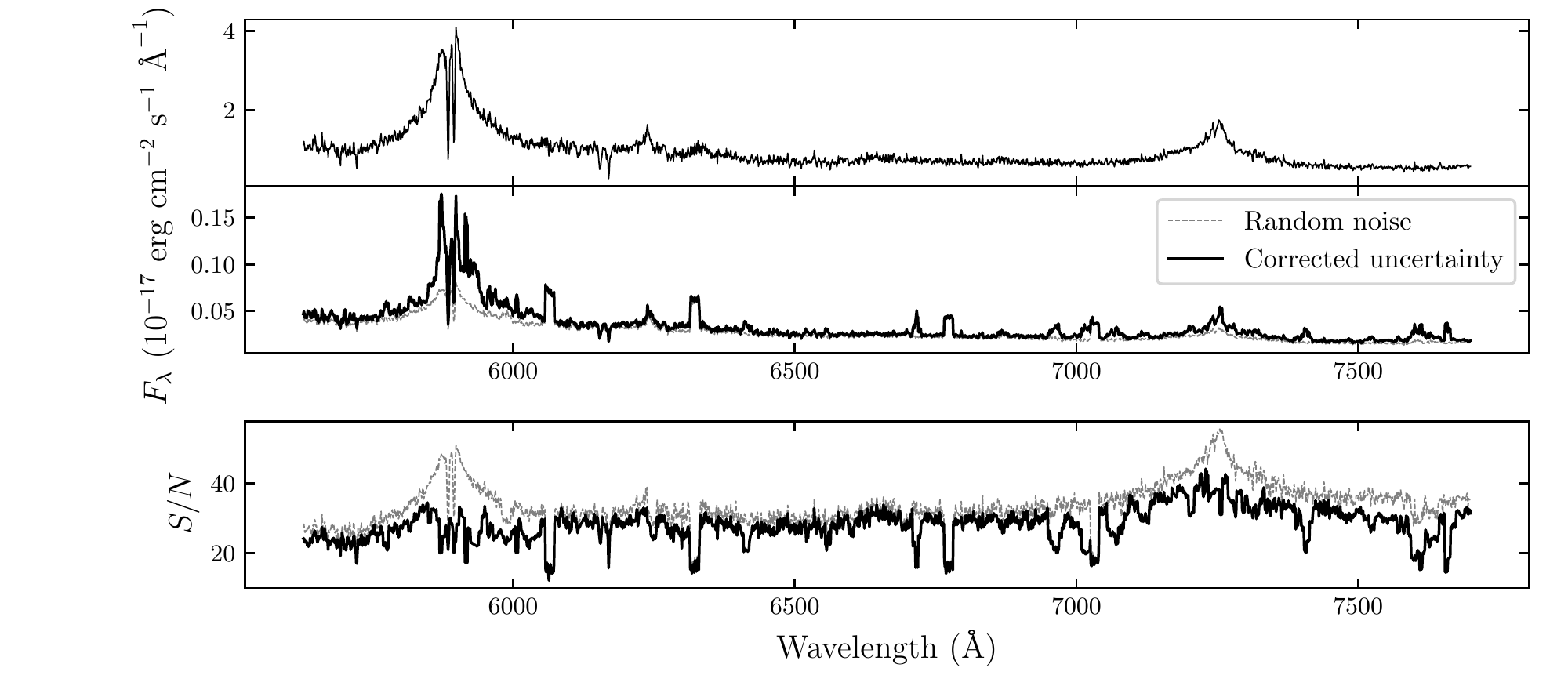}  
\caption{\textit{Upper panel}: Spectrum of QSO image A for the first epoch. \textit{Middle panel}: Original random uncertainty (dotted line) and the corrected uncertainty (solid line) including calibration uncertainty and other additional sources of noise. \textit{Bottom panel}: Original $S/N$ calculated using only random uncertainty (dotted line) and the $S/N$ with the corrected uncertainty (solid line).
\label{fig:qso_spec_uncertainty}}
\end{center}
\end{figure*}

\subsection{Correcting the uncertainties for all spectra}
With the results of Section \ref{sect:calib_uncertainty}, we can now correct the uncertainties on all of the spectra to include the additional uncertainty due to flux calibration.
For a general spectrum from epoch $i$, this is a simple correction:
\begin{align}
\sigma_{q,i,tot} = \sqrt{\sigma_{q,i}^2 + \left(\frac{\sigma_{f,0}}{f_i}\right)^2 \cdot q_i^2}.
\label{eq:uncertainty_correction}
\end{align}
Here, $f_i$, is the spectrum of the standard star used to determine the calibration uncertainty, $\sigma_{f,0}$ is the inferred calibration uncertainty spectrum, $q_i$ is the science spectrum, and $\sigma_{q,i}$ is the random uncertainty for that science spectrum.

Since we had to bin the standard star spectrum in order to make a measurement of $\sigma_{f,0}$, we do not have a measurement at every pixel in the science spectra.
To make the correction, we first linearly interpolate $\sigma_{f,0}$ back onto the wavelength scale of the science spectra and then calculate the uncertainties at every wavelength pixel according to Equation \ref{eq:uncertainty_correction}.
The spectrum of quasar image A for the first epoch with its adjusted total uncertainty is shown in Figure \ref{fig:qso_spec_uncertainty}.


\section{Fitting the spectra}
\label{sect:spectral_decomp}
Measuring the black hole mass using reverberation mapping methods requires an accurate measurement of the broad emission line flux and shape.
Contaminating emission or absorption lines in the vicinity of the broad emission line of interest can lead to over- or under-estimates of the broad emission line flux in different parts of the line profile.
It is important, then, to disentangle all components of the spectra and isolate, e.g., the \civ\ emission line.

To do so, we construct a model spectrum that consists of several components, described below.
We then fit that model to the data using the {\sc Python} programs \texttt{scipy.optimize.minimize} and \texttt{emcee}.
We first use \texttt{scipy.optimize.minimize} to find maximum likelihood fit and then use \texttt{emcee} to explore the parameter space, allowing us to assign uncertainties to our fits.

\subsection{Components}
\label{sect:components}

\begin{figure*}[t!]
\begin{center}
\includegraphics[width=7.0in]{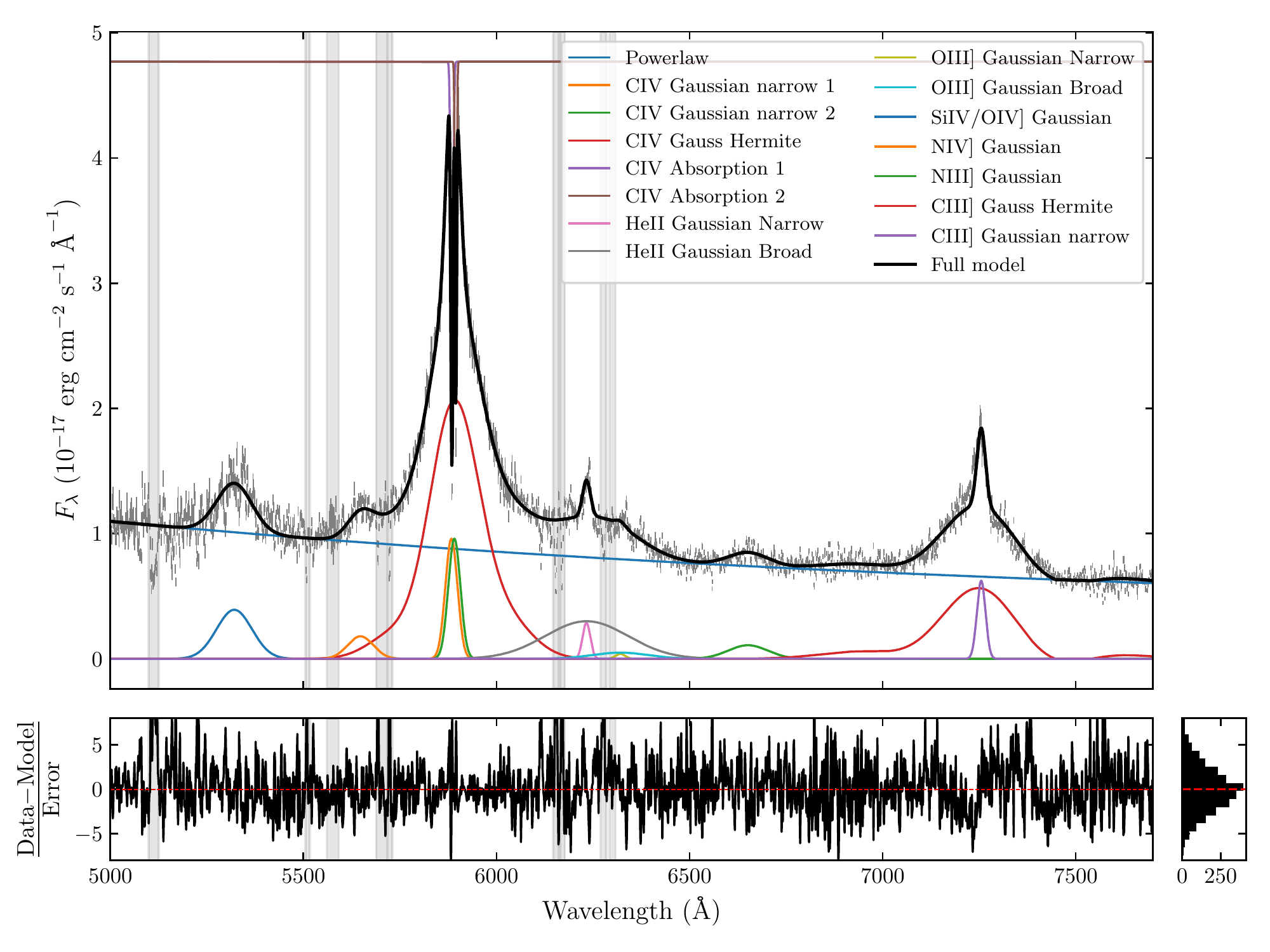}  
\caption{
Example model fit to the quasar image A spectrum from 2017 June 21.
The top panel shows the data in grey, the total model fit in black, and the individual components in color.
Greyed out bands indicate wavelength ranges that were masked out during the fitting procedure either due to bad pixels or un-modeled absorption features.
The normalized residuals are shown in the bottom panel.
\label{fig:spectral_decomp}}
\end{center}
\end{figure*}

\begin{itemize}

\item {\it AGN featureless continuum}:
We model the AGN featureless continuum using a power law, $F_\lambda = C_{\rm 5000} \lambda^\alpha$, where $C_{\rm 5000}$ is the normalization at 5000\AA, observed frame, and $\alpha$ is the spectral index.
Both of these parameters are free to vary for every epoch.

\item \civdub:
\civ\ is modeled as a sum of a fourth order Gauss-Hermite function and two narrower Gaussians at the doublet central wavelengths.
We initially attempted to model the line using only the Gauss-Hermite component, but were unable to simultaneously fit the broad wings and the strong core of the line.
The combination of the three functions allows the Gauss-Hermite function to fit the broad line profile and the narrower Gaussian component fits the core of the line.
After fitting to the mean spectrum, all Gauss-Hermite parameters were allowed to vary while the narrow Gaussian components were kept fixed for each epoch.

\item {\it Narrow absorption on} \civdub: 
There are two strong absorption features near the peak of the \civ\ emission line.
Since masking these features removes important information needed to adequately fit the emission line, we model the absorption with two narrow Gaussians with amplitudes ranging from $0$ to $-1$ and sharing a common width.
After summing all other components, we apply the absorption by multiplying the model by $1 - A_1 \mathcal{N}(\lambda_1,\sigma^2) - A_2 \mathcal{N}(\lambda_2,\sigma^2)$, where $A_i$ are the amplitudes, $\lambda_i$ are the central wavelengths, and $\sigma$ is the shared standard deviation of the Gaussians.
After fitting to the mean spectrum, these components are kept fixed for each epoch.

\item \ciii\ $\lambda$1909:
\ciii\ is modeled with a fourth order Gauss-Hermite function plus a narrow Gaussian at the central wavelength.
All Gauss-Hermite parameters are allowed to vary for each epoch, while the narrow Gaussian components are kept fixed.

\item \siivoiv\ $\lambda$1400: 
The \siivoiv\ blend is modeled using a single broad Gaussian.
Once fit to the mean spectrum, the amplitude and width are allowed to vary for each epoch, but the centroid is fixed.

\item \oiiisemi\ $\lambda$1663:
The \oiiisemi\ line is modeled with a narrow and broad Gaussian component.
The broad component parameters are allowed to vary for each epoch while the narrow component is fixed.

\item \heii\ $\lambda$1640:
The \heii\ line is modeled with a narrow and broad Gaussian component.
The broad component parameters are allowed to vary for each epoch while the narrow component is fixed.

\item \niv\ $\lambda$1486:
The \niv\ emission line, located in the blue wing of \civ, is modeled using a broad Gaussian.
All parameters are allowed to vary for each epoch.

\begin{figure*}[t!]
\begin{center}
\includegraphics[width=7in]{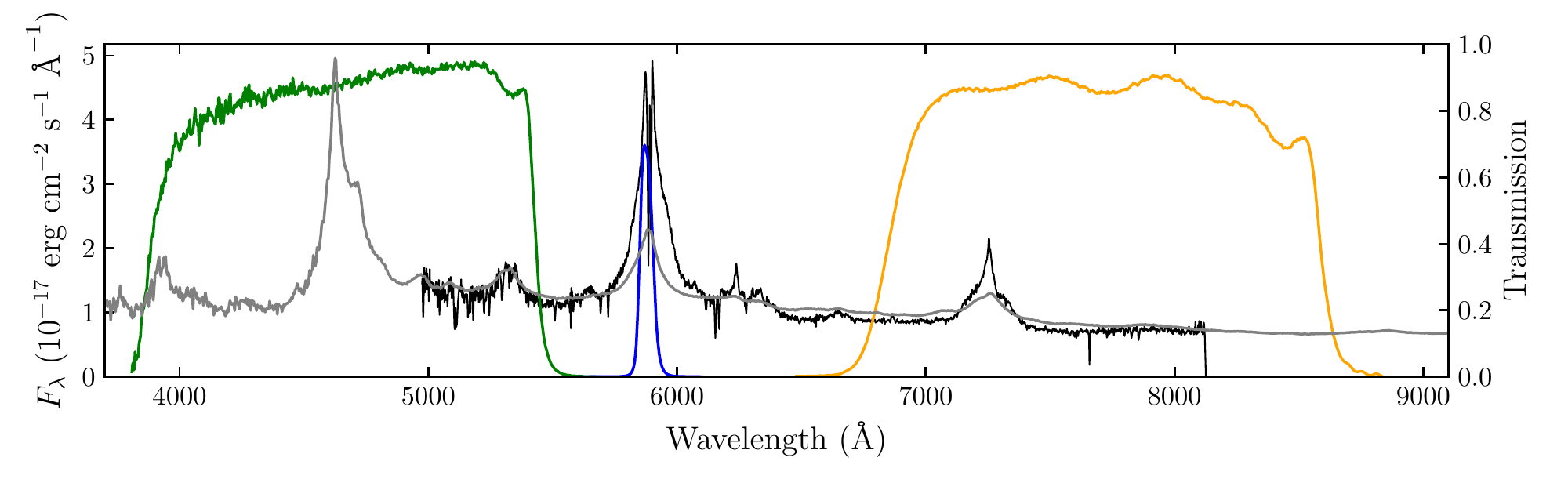}  
\caption{
Broad-band and narrow-band filter transmission curves shown in relation to the SDSS J2222+2745 mean quasar spectrum (black) and the \citet{vandenberk++01} mean quasar spectrum (grey).
The green, orange, and blue curves are the $g$-band, $i$-band, and narrow-band filter transmission curves, respectively.
Note that the \citet{vandenberk++01} spectrum is plotted with an arbitrary scale and is only meant to show the positions of the emission lines in relation to the filters.
\label{fig:filters}}
\end{center}
\end{figure*}

\item \niii\ $\lambda$1750:
The broad \niii\ emission line lies between the \civ\ and \ciii\ emission lines and is modeled as a single Gaussian.
We are able to fit this component in the mean spectra of each overving season, but the $S/N$ of the individual spectra is not high enough to constrain the fit.
We therefore keep this component fixed when fitting each epoch.

\end{itemize}

\subsection{Fitting procedure}

Before fitting the spectra for each individual epoch, we fit the mean quasar spectrum of each image to provide the initial parameter guesses.
We start with the mean spectrum of image A which is the brightest of the three images and thus has the highest signal-to-noise ratio.
Initially, we mask all of the emission and absorption features and fit the featureless power law continuum to the unmasked regions.
We then unmask the \civ\ emission line and fit the Gauss-Hermite function, narrower Gaussians, and two absorption features, keeping the power law parameters and absorption trough widths fixed.
Allowing the absorption width to vary at this stage typically led to one of the absorption features encompassing both absorption troughs and the fits failed to converge.
Next, we keep the Gauss-Hermite and power law components fixed and allow all parameters in the narrow \civ\ Gaussians and two absorption features to vary.
Finally, we allow all of the model parameters to vary.

We then unmask the emission features near \civ\ and add the \oiiisemi, \heii, \siivoiv, \niii, and \niv\ components to the model.
We keep the parameters of the power law, \civ, and absorption components fixed and allow all parameters of the new components to vary, and then repeat the fit allowing all parameters to vary.
Finally, we unmask the \ciii\ broad emission line, introduce the \ciii\ Gauss-Hermite and Gaussian components to the model, and fit the entire model, allowing all parameters to vary.
This becomes the best-fit model for the quasar image A mean spectrum.

Next, we fit the image B and image C mean spectra in two steps.
Starting with the best-fit model for the image A spectrum, we allow the power law index, power law normalization, and the amplitudes of all other components to vary while keeping everything else fixed.
We then do the fit again, keeping only the emission line centroids and absorption line centroids and widths fixed.
This provides the best-fit model for the mean spectra of images B and C.

Using the best-fit models for the three mean spectra as the starting guesses, we then proceed to fit the per-season mean spectra.
Finally, we use the resulting fits as the starting guesses to fit the spectra for each individual epoch.
For this stage, all broad components are allowed to vary except \niii\ (see Section \ref{sect:components}), and all narrow components are kept fixed.
An example fit to one of the quasar spectra is shown in Figure \ref{fig:spectral_decomp}.

\section{Broad-band and narrow-band photometry}
\label{sect:photometry}
Throughout the spectroscopy campaign, we also obtained narrow-band and broad-band photometry of the three quasar images at roughly twice the spectroscopic cadence.
The higher-cadence photometry provides $g$- and $i$-band continuum light curves as well as narrow-band \civ\ measurements to supplement the spectroscopic observations.
The transmission curves of the three filters are shown in Figure \ref{fig:filters}, plotted on the quasar mean spectrum.

\begin{figure}[h!]
\begin{center}
\includegraphics[width=3in]{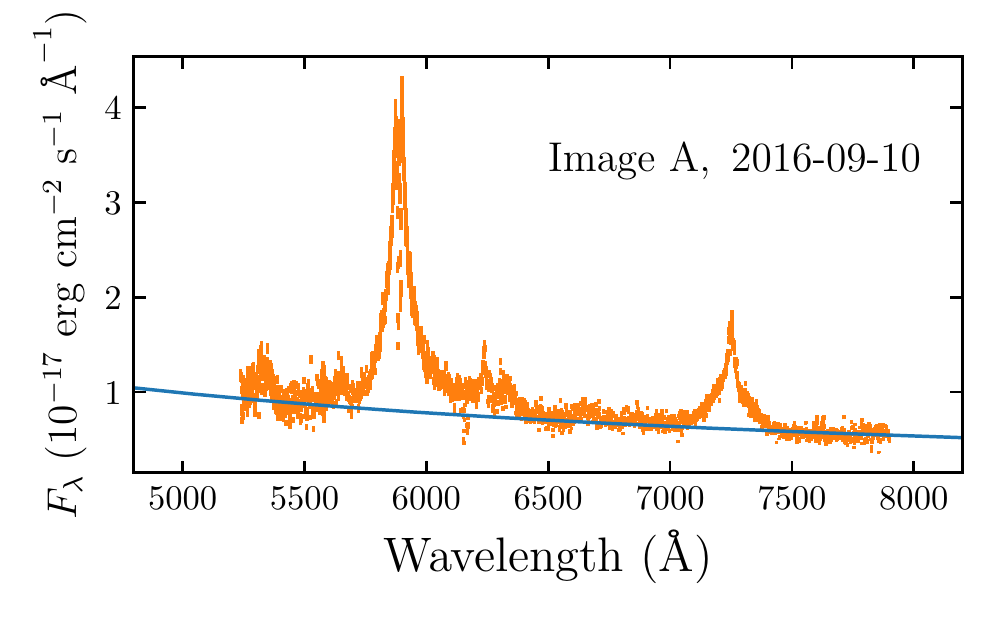}  
\includegraphics[width=3in]{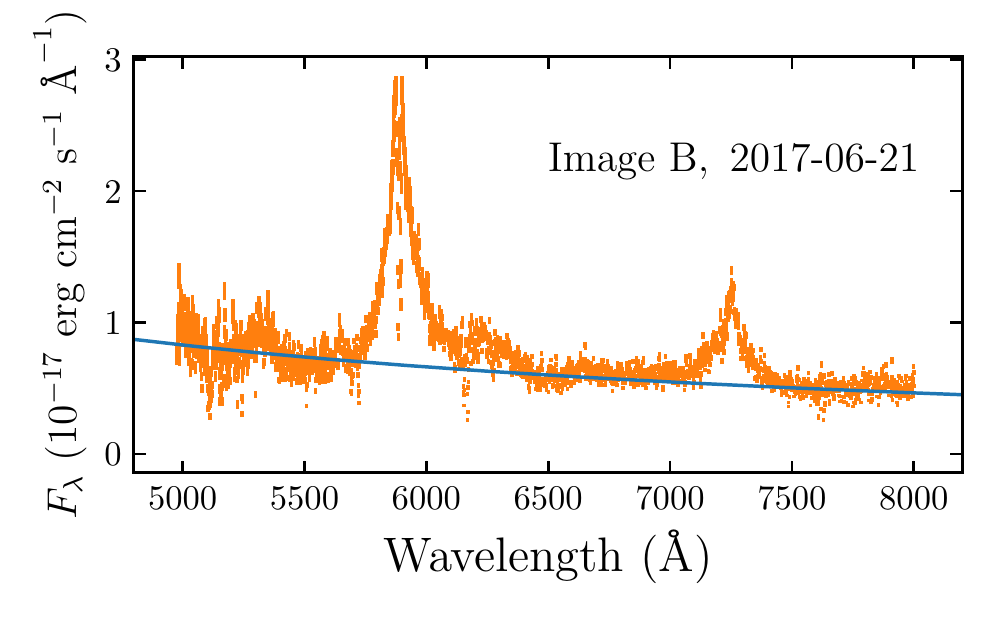}  
\includegraphics[width=3in]{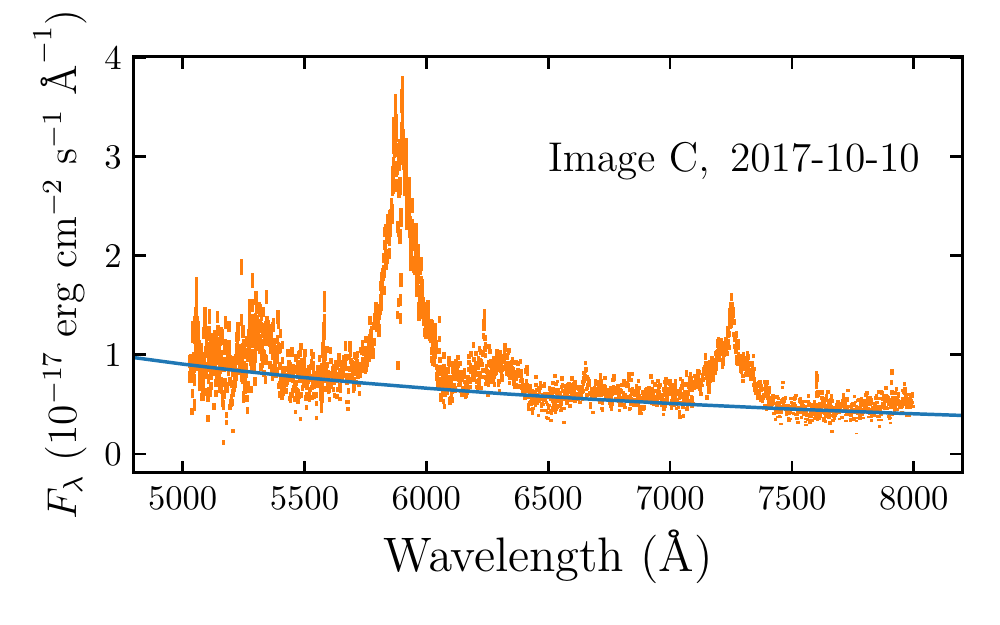}  
\caption{Comparison of the power law determined {\it only} from broad-band photometry (blue) to a spectrum taken on the same day (orange). The examples shown are, from top to bottom: Image A, 2016 September 9; Image B, 2017 June 21; Image C, 2017 October 10.
\label{fig:photometry_powerlaw_spectra_comparison}}
\end{center}
\end{figure}

The broad-band filters and narrow-band filter provide reasonable estimates of the continuum and \civ\ line flux, respectively, but these measurements are imperfect since the broad-band filters cover emission line regions and the narrow-band filter includes both the \civ\ emission line and the power law continuum.
To disentangle these components, we can examine nights in which we obtained both spectroscopy and photometry and produce conversion factors between the three measurements.
The spectral range of the data does not extend all the way to the ends of the broad-band filters, so we utilize the mean quasar spectrum from \citet{vandenberk++01}, also shown in Figure \ref{fig:filters}.

Since the \citet{vandenberk++01} spectrum consists of emission lines plus a power law continuum that are not of equal strength to the quasar spectrum, we need to rescale the \citet{vandenberk++01} spectrum.
We first subtract a power law continuum fit from the \citet{vandenberk++01} spectrum so that we are left with only the emission line component.
We then multiply by a scale factor to match the \civ\ emission line strength of the SDSS J2222+2745 mean spectrum.
Finally, we add the power law continuum component of the SDSS J2222+2745 mean spectrum, leaving us with an extended estimate of the SDSS J2222+2745 spectrum.
Here, we are assuming that the line ratios from the \citet{vandenberk++01} spectrum are the same as the line ratios of the quasar mean spectrum, and we do not allow for any variations in these values.

Next we integrate the power law continuum component as well as the full rescaled \citet{vandenberk++01} spectrum over the $g$- and $i$-band transmission curves to determine the continuum contribution to the flux in the two bands.
Doing so, we find that $89\%$ of the flux in the $g$ band and $85\%$ of the flux in the $i$ band comes from the continuum.
We perform the same procedure using the mean spectra for each individual season and find that the continuum fractions do not change by more than $5\%$ throughout the campaign.

Using these percentages, we scale the measured broad-band photometry to obtain the continuum fluxes in these bands.
We then use these two values to determine the power law index and normalization for each photometric epoch.
As a check on the method, the power law determinations from the $g$- and $i$-band photometry can be compared with spectra on days where the spectroscopic and photometric observations overlap.
Three example fits are shown in Figure \ref{fig:photometry_powerlaw_spectra_comparison}.

Finally, we integrate the power law continuum fit over the narrow-band filter transmission curve to determine the continuum contribution in this band for every epoch.
Subtracting this from the narrow-band measurements, we are left with the \civ\ contribution to these fluxes.

\section{Results and Discussion}

\begin{figure*}[h!]
\begin{center}
\includegraphics[width=\textwidth]{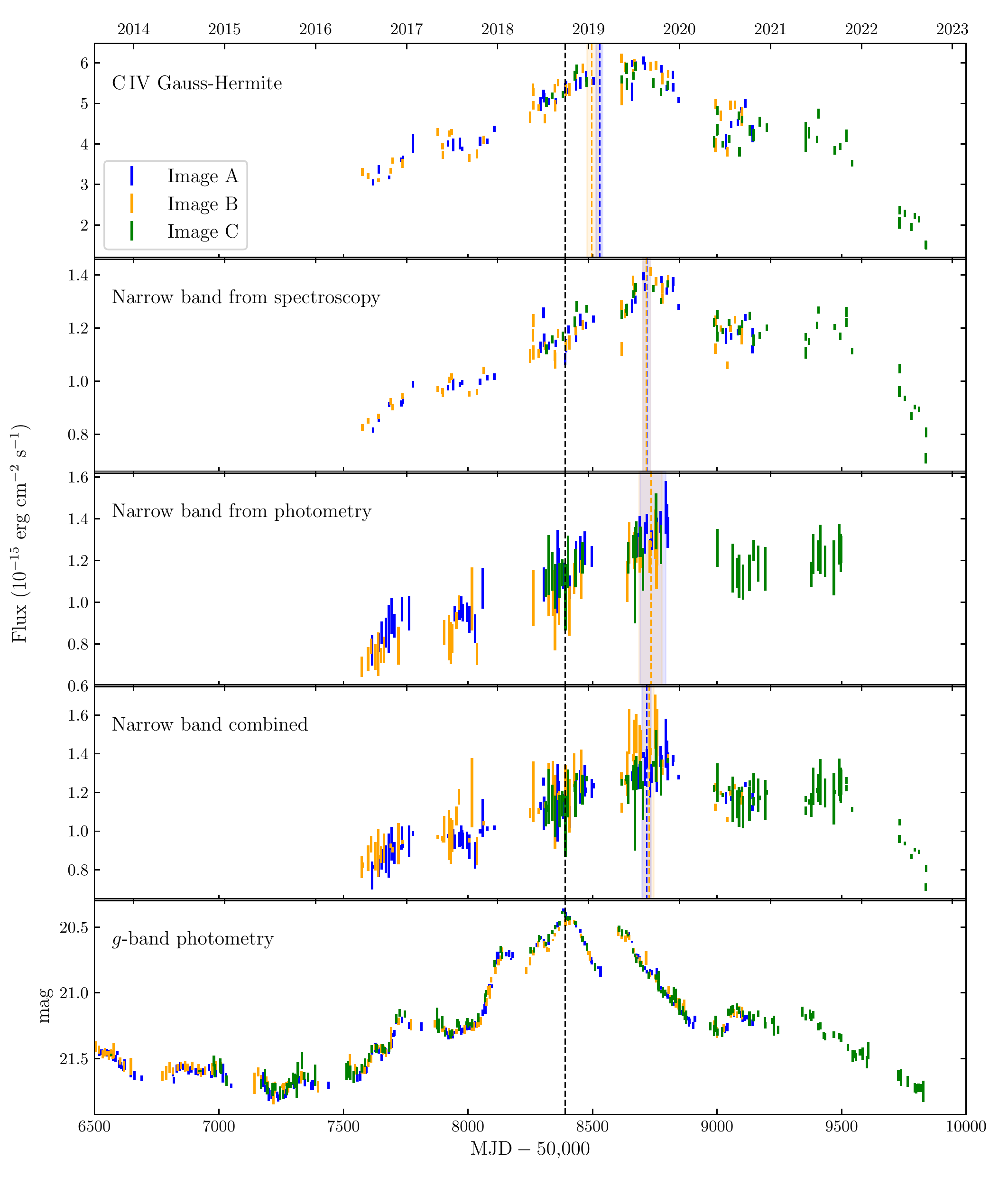}  
\caption{
\civ\ emission line light curves and $g$-band continuum light curve. The blue, orange, and green points correspond to quasar images A, B, and C, respectively.
All data points have been shifted in time according to the measured gravitational lensing time delays and scaled according to the measured relative magnifications.
The vertical dashed black line marks the peak of the $g$-band light curve at ${\rm MJD}-50{,}000 = 8390$.
The vertical blue and orange dashed lines and shaded bands show the $\tau_{\rm cen}$ and $\tau_{\rm peak}$ lags (Section \ref{sect:lags}), respectively, shifted relative to ${\rm MJD}-50{,}000 = 8390$.
\label{fig:light_curve_gh_narrow}}
\end{center}
\end{figure*}

\begin{figure*}[h!]
\begin{center}
\includegraphics[width=\textwidth]{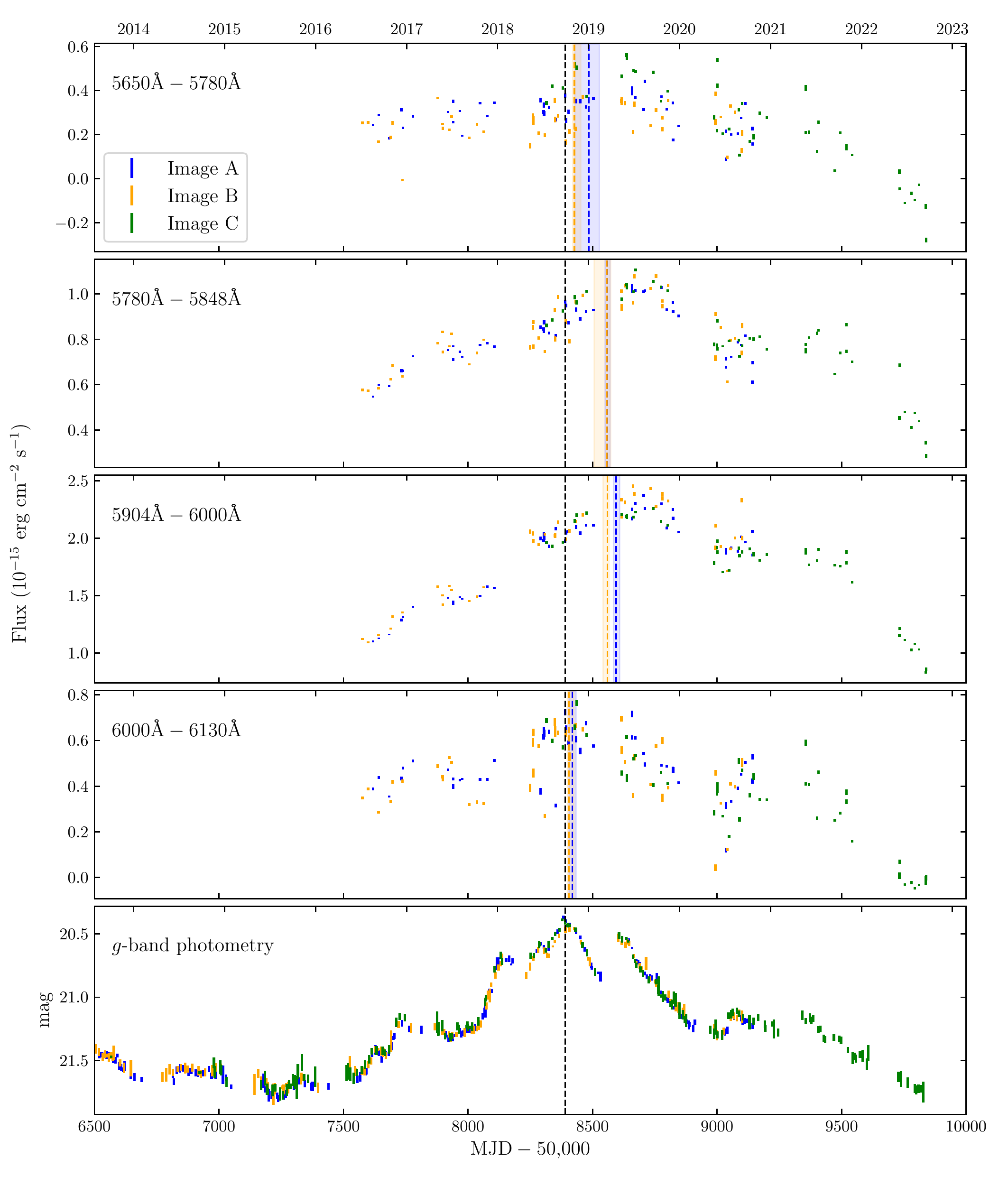}  
\caption{Same as Figure \ref{fig:light_curve_gh_narrow}, but showing the four wavelength windows.
\label{fig:light_curve_windows}}
\end{center}
\end{figure*}

\subsection{Light curves}
Using the model fits to the data from Section \ref{sect:spectral_decomp}, we measure emission line fluxes in a number of ways.
First, we take the Gauss-Hermite fits to the spectra which represent the broad \civ\ emission component.
For each spectrum, we take the sample of fits from the MCMC chains and compute the integrated \civ\ Gauss-Hermite flux for each.
From these values, we compute the median integrated flux value and 68\% confidence interval.
The benefit of this approach is that we can calculate the full contribution of broad \civ\ despite the strong absorption in the core, all while maintaining realistic uncertainty estimates.
While we do lose the information contained in the residuals of the emission line fits, integrating over the full emission line profile likely averages this information out.
The resulting light curve is shown in the top panel of Figure \ref{fig:light_curve_gh_narrow}.
Each data point is shifted in time using the measured gravitational time delays of $\Delta\tau_{AB} = -42.44$ days and $\Delta\tau_{AC} = 696.65$ days (H. Dahle 2020, private communication), setting image B as the reference.
The fluxes are then scaled using the measured magnitude offsets from gravitational lensing of $\Delta m_{AB} = 0.353$ and $\Delta m_{AC} = 0.515$ (H. Dahle 2020, private communication).

In addition, we find evidence for microlensing in image B, which affects the compact continuum source, but not the extended broad line region.
Since the magnifications were measured using the $g$-band continuum, which \textit{is} affected by microlensing, we need to apply a correction factor of $\Delta m = 0.072$ to all emission line measurements on image B.
This correction is discussed further in Section \ref{sect:microlensing}.

Next, we take the full model fits to the spectra and subtract all model components {\it except} the \civ\ emission and absorption components.
This procedure leaves the emission line data for \civ, but with the contaminating components in the emission line wings removed and the continuum subtracted.
Using this method, we retain the information contained in the residuals of the spectral fits and can directly compare the measurements to the narrow-band photometry, which also includes the narrow \civ\ emission and absorption features.
With these spectra, we compute the \civ\ emission line flux for each epoch integrated over four wavelength windows: 5650\AA$-$5780\AA\ ($-12{,}420$ to $-5800~{\rm km~s}^{-1}$), 5780\AA$-$5848\AA\ ($-5800$ to $-2340~{\rm km~s}^{-1}$), 5904\AA$-$6000\AA\ ($510$ to $5400~{\rm km~s}^{-1}$), and 6000\AA$-$6130\AA\ ($5400$ to $12{,}020~{\rm km~s}^{-1}$) and the narrow-band transmission curve, the results of which are shown in Figures \ref{fig:light_curve_gh_narrow} and \ref{fig:light_curve_windows}.

Finally, we include the measurements from the narrow-band photometry to supplement the light curve measured from the spectra.
Using the procedure described in Section \ref{sect:photometry}, we use the $g$- and $i$-band photometry to calculate the continuum contribution in the narrow-band filter.
We then subtract these values from the measured narrow-band photometry to produce the light curve shown in Figure \ref{fig:light_curve_gh_narrow}, panel 3.
The combined narrow-band light curve from spectroscopy and photometry is shown in panel 4.

\subsection{Emission line time lags}
\label{sect:lags}
We use cross-correlation analysis to measure the lag between the continuum and emission line fluctuations and the flux randomization and random subset selection method \citep[FR/RSS;][]{peterson98} to estimate uncertainties.
We compute the lags for each emission line light curve compared to the $g$-band continuum from photometry.

\begin{figure*}[h!]
\begin{center}
\includegraphics[width=\textwidth]{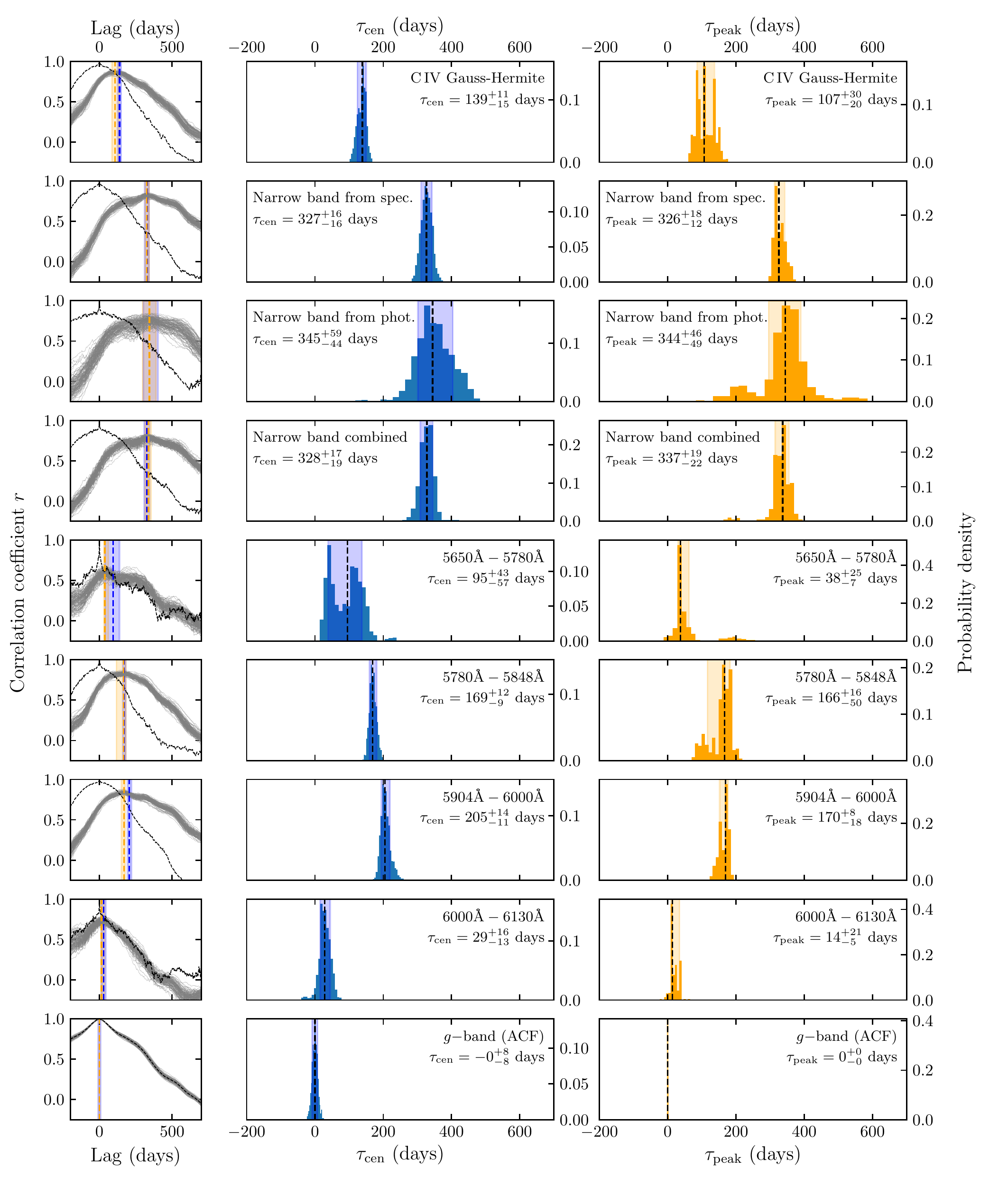}  
\caption{\textit{Left:} Distribution of ICCFs calculated using the FR/RSS method (grey). The vertical dashed blue and orange lines show the median $\tau_{\rm cen}$ and $\tau_{\rm peak}$ values, respectively, and the shaded regions show the 68\% confidence intervals. The black dashed line is the auto-correlation function (ACF) for comparison. \textit{Middle:} Distribution of $\tau_{\rm cen}$ values. The median and 68\% confidence interval are indicated by the vertical dashed line and shaded region. \textit{Right:} Same as the middle panel, but for $\tau_{\rm peak}$. In the bottom panel, we show the $g$-band ACF. Note that all lags here are in the observed frame.
\label{fig:ccfs}}
\end{center}
\end{figure*}

For each light curve, we produce 1000 sample light curves by taking a random subset of the observed light curve and shifting the fluxes randomly according to their uncertainties.
We then compute the interpolated cross-correlation function \citep[ICCF;][]{Gaskell+87, White+94} for every light curve in the sample and compute the lag $\tau_{\rm peak}$ at which the ICCF reaches its peak value $r_{\rm max}$, and the lag centroid defined by
\begin{align}
\tau_{\rm cen} = \frac{\int \tau r(\tau)d\tau}{\int r(\tau) d\tau}
\end{align}
for $r(\tau) \ge 0.8 r_{\rm max}$, where $r(\tau)$ is the correlation coefficient.
The ICCFs and lag distributions are shown in Figure \ref{fig:ccfs}.
We then calculate the median values and 68\% confidence intervals for $\tau_{\rm cen}$ and $\tau_{\rm peak}$, which are reported in Table \ref{table:lags}.

\begin{figure*}[t!]
\begin{center}
\includegraphics[width=\textwidth]{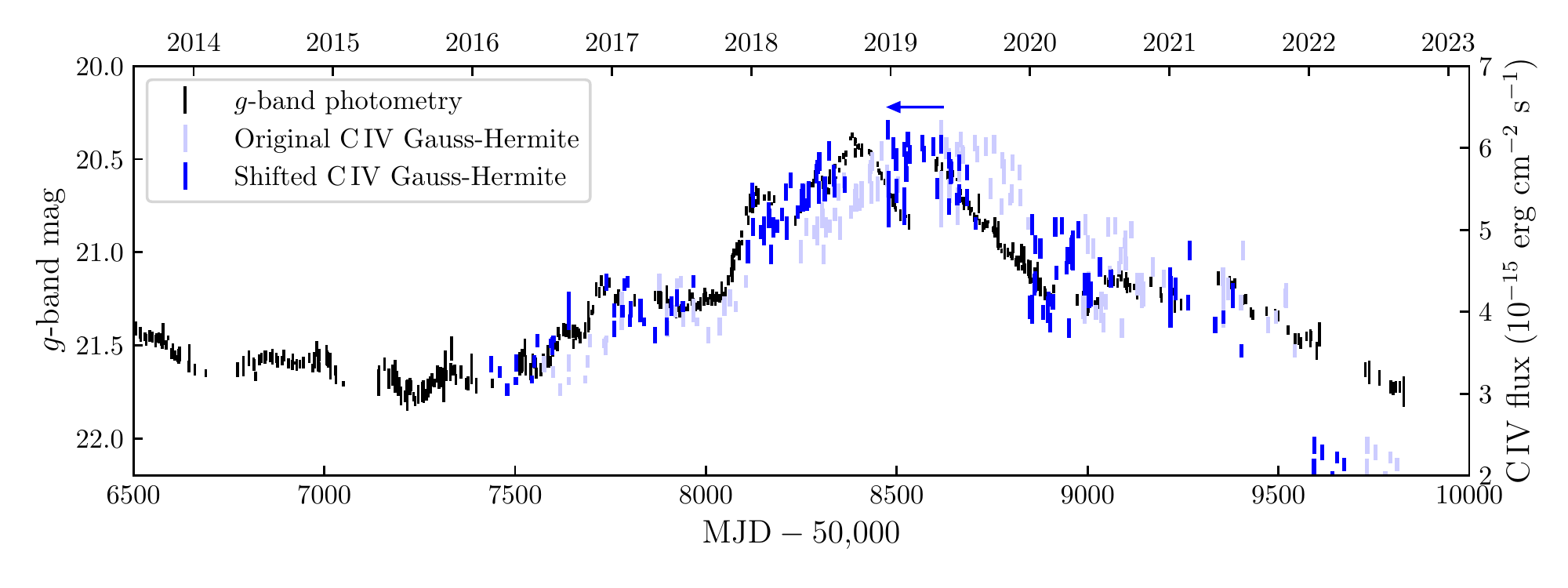}  
\caption{
Light curves for the $g$-band photometry and \civ\ Gauss-Hermite component.
The \civ\ light curve has been shifted by the measured time delay $\tau_{\rm cen} = 139$ days (observed frame) to align with the $g$-band continuum light curve.
\label{fig:light_curve_overlay}}
\end{center}
\end{figure*}

In Figure \ref{fig:light_curve_overlay}, we show the light curve for the \civ\ Gauss-Hermite emission component plotted alongside the $g$-band continuum light curve.
Also shown is the \civ\ Gauss-Hermite light curve shifted in time by the measured time lag of $\tau_{\rm cen} = 36.5$ days (converted to the observed frame).
The emission line light curve clearly tracks the fluctuations of the $g$-band continuum until ${\rm MJD}-50{,}000 \approx 9600$.
Additional future monitoring data are necessary to shed light on the reason for the decrease in flux beyond this point.

\begin{deluxetable}{lcc}
\tablecaption{Rest frame emission line lags}
\tablewidth{0pt}
\tablehead{ 
\colhead{Light curve} & 
\colhead{$\tau_{\rm cen}$ (days)} & 
\colhead{$\tau_{\rm peak}$ (days)}
}
\startdata
\civ\ Gauss-Hermite & $36.5^{+2.9}_{-3.9}$ & $28.1^{+8.0}_{-5.4}$ \\
Narrow~band~from~spec. & $85.9^{+4.2}_{-4.2}$ & $85.5^{+4.6}_{-3.3}$ \\
Narrow~band~from~phot. & $91^{+16}_{-12}$ & $91^{+12}_{-13}$ \\
Narrow~band~combined & $86.2^{+4.5}_{-5.0}$ & $88.6^{+4.9}_{-5.8}$ \\
\makecell[l]{5650\AA$-$5780\AA \\ ($-12{,}420$ to $-5800~{\rm km~s}^{-1}$)} & $25^{+11}_{-15}$ & $9.9^{+6.6}_{-1.8}$ \\
\makecell[l]{5780\AA$-$5848\AA \\ ($-5800$ to $-2340~{\rm km~s}^{-1}$)} & $44.4^{+3.1}_{-2.4}$ & $43.8^{+4.2}_{-13.2}$ \\
\makecell[l]{5904\AA$-$6000\AA \\ ($510$ to $5400~{\rm km~s}^{-1}$)} & $53.9^{+3.7}_{-2.9}$ & $44.5^{+2.0}_{-4.6}$ \\
\makecell[l]{6000\AA$-$6130\AA \\ ($5400$ to $12{,}020~{\rm km~s}^{-1}$)} & $7.5^{+4.2}_{-3.5}$ & $3.7^{+5.5}_{-1.2}$
\enddata
\tablecomments{Rest frame emission line lags measured using cross-correlation analysis. 
\label{table:lags}}
\end{deluxetable} 

In most cases, both $\tau_{\rm cen}$ and $\tau_{\rm peak}$ are in close agreement with each other.
Notable exceptions are the measurements for the bluest wavelength bin, although they are consistent to within the uncertainties.
Examining Figure \ref{fig:ccfs}, it is clear that the disagreement is an artifact of the double-peaked nature of the distribution of ICCF centroids, and the lag measurements for this light curve have large uncertainties as a result.
This is unsurprising given that this light curve is for a low $S/N$ part of the spectrum.

\begin{figure}[h!]
\begin{center}
\includegraphics[width=3.0in]{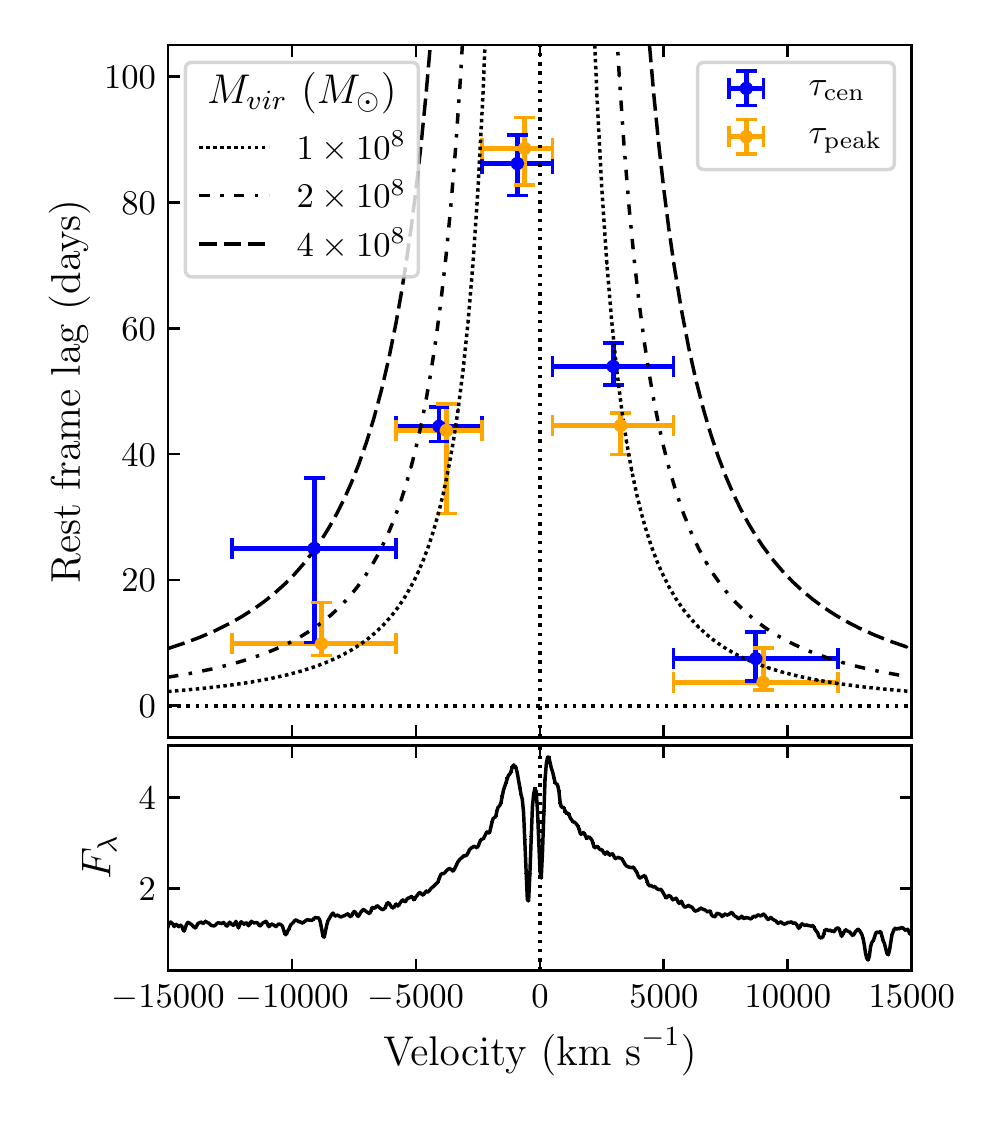}  
\caption{\textit{Top:} Rest frame emission line lag measurements, $\tau_{\rm cen}$ and $\tau_{\rm peak}$, for the four wavelength windows and combined narrow-band filter light curves (central bin).
The $x$-axis error bars indicate the wavelength ranges of the windows and the $y$-axis error bars are the 1$\sigma$ uncertainties on the lag measurements.
Since the middle bin uses the narrow-band filter, we simply show the error bars extended to the edges of the neighboring windows.
We shift the $\tau_{\rm peak}$ measurements by $+300~{\rm km~s}^{-1}$ for visibility purposes.
Virial envelopes are overplotted for three choices of $M_{\rm vir} = M_{\rm BH}/f$.
\textit{Bottom: } Mean quasar image A \civ\ profile. The $y$-axis units are $10^{-17}~{\rm erg~s}^{-1}~{\rm cm}^{-2}~{\rm \AA}^{-1}$.
\label{fig:lags_velocityresolved}}
\end{center}
\end{figure}

In Figure \ref{fig:lags_velocityresolved}, we show the lag measurements for the four velocity bins and within the narrow-band filter.
There is clear velocity structure in the line profile, with the core of the line responding on longer time scales and the wings responding on shorter time scales.
This behavior is expected if the emission near the core of the line is produced by gas that is farther from the BLR center and the wing emission is produced by the high-velocity gas closer to the center.
The symmetry of the velocity-lag structure suggests that the \civ-emitting gas motions are dominated by bound circular orbits rather than inflowing or outflowing motions \citep[see, e.g.,][Figure 10]{bentz09}.
We also show virial envelopes defined by $v^2 = GM_{\rm vir}/c\tau$ for three choices of $M_{\rm vir}$: $1\times 10^8$, $2\times 10^8$, and $4\times 10^8$ M$_\odot$.

\subsection{Radius-luminosity relation}
\label{sect:rL}
\begin{figure*}[h!]
\begin{center}
\includegraphics[width=7in]{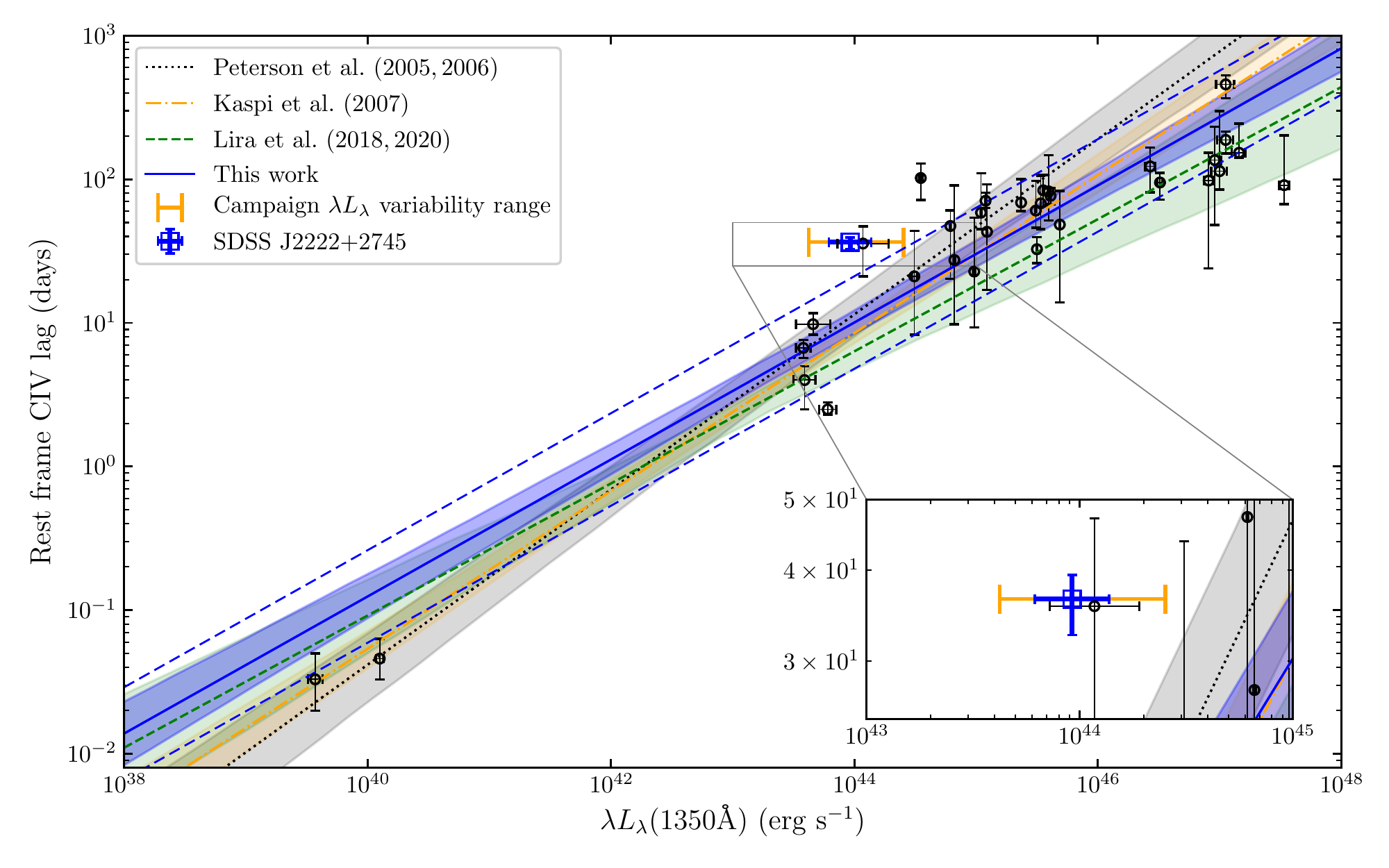}  
\caption{\civ\ BLR radius luminosity relation. Black circles with error bars represent previous measurements in the literature, and the blue square is SDSS J2222+2745.
The orange bars show the range over which SDSS J2222+2745 fluctuated in $\lambda L_\lambda(1350{\rm \AA})$ over the duration of the campaign.
The four diagonal lines with shaded bands are the median and 68\% confidence intervals of the fits to the $r_{\rm BLR}-L$ relation presented by \citet{Peterson++05}, \citet{Kaspi++07}, \citet{Lira++18}, and this work.
The confidence interval bands are calculated by producing 1000 Monte Carlo line iterations based on the reported fits, assuming Gaussian uncertainties, and computing the 68\% confidence interval at each $\lambda L_\lambda$ value.
The dashed blue lines indicate the intrinsic scatter for the fit from this work and are offset by $\epsilon = 0.32$ dex from the best fit line.
We show a zoomed-in inset of the region around the SDSS J2222+2745 data point to show more detail.
\label{fig:rL}}
\end{center}
\end{figure*}

Using the measured lags, we can place SDSS J2222+2745 on the BLR radius-luminosity relationship for \civ.
Since the quasar luminosity varied significantly over the duration of the campaign ($\sim$1.5 magnitudes in $g$-band), the placement on the $r_{\rm BLR}-L$ relation depends on when the luminosity measurement is made.
Additionally, the luminosity calculation depends on the intrinsic magnification of the images due to gravitational lensing.
We use the mean quasar spectrum for each image with the magnifications determined by \citet{Sharon++17} through lens modeling to measure the mean $\lambda L_\lambda(1300{\rm \AA})$.
Taking the average of the measurements for each quasar image, we find $\log_{10}[\lambda L_\lambda(1350{\rm \AA}) / {\rm erg~s}^{-1}] = 43.96 \pm 0.18$.
Next, we take the minimum and maximum luminosities reached over the duration of the campaign to find that the quasar ranged from $\log_{10}[\lambda L_\lambda(1350{\rm \AA}) / {\rm erg~s}^{-1}] = 43.62$ to $44.40$.
We show the results for $\tau_{\rm cen}$ in Figure \ref{fig:rL} along with previous measurements by \citet[][7 points]{Peterson++05, Peterson++06}, \citet[][1 point]{Kaspi++07}, \citet[][6 points]{Lira++18, Lira++20}, \citet[][2 points]{Hoormann++19}, \citet[][16 points, ``gold sample'' only, ICCF lags]{Grier++19}.
We omit two data points from \citet[][CT250 and CT320]{Lira++18} since either the median or $1\sigma$ uncertainties include negative lags.

Using these data, we fit a linear regression of the form
\begin{align}
\log_{10}\left(\frac{\tau_{\rm CIV}}{{\rm days}}\right) = \alpha + \beta \log_{10}\left[\frac{\lambda L_\lambda(1350{\rm \AA})}{10^{44}~{\rm erg~s}^{-1}}\right] + \mathcal{N}(0,\epsilon^2),
\end{align}
$\mathcal{N}(0,\epsilon^2)$ is a normal distribution with mean 0 and variance $\epsilon^2$, and $\epsilon$ is the intrinsic scatter.
We use the {\sc IDL} routine \texttt{linmix\_err} \citep{linmix} which assumes Gaussian uncertainties, so we take the mean of the upper and lower uncertainties when they are not equal.
We find a slope $\alpha = 1.00 \pm 0.08$, intercept $\beta = 0.48 \pm 0.04$, and intrinsic scatter $\epsilon = 0.32 \pm 0.06$.
The fit is plotted in Figure \ref{fig:rL} along with the three fits by \citet{Peterson++05}, \citet{Kaspi++07}, and \citet{Lira++18}.
We do not include the fits by \citet{Hoormann++19} and \citet{Grier++19} since they were calculated before the lag corrections were presented by \citet{Lira++20}.
SDSS J2222+2745 lies 0.58 dex above the mean $r_{\rm BLR}-L$ relation, but this offset can be explained by the large intrinsic scatter as well as the large variability in $\lambda L_\lambda$ over the duration of the campaign.

We should note that the lag measurement for SDSS J2222+2745 is extremely well secured due to the combination of high quasar variability, high cadence monitoring, and high $S/N$ data.
However, the fit to the $r-L$ relation is primarily driven by the sheer quantity of data points from large-scale campaigns such as that described by \citet{Grier++19}, even with the bigger uncertainties.
While these types of campaigns are certainly beneficial, future fits to the $r-L$ relation would benefit greatly from more high-precision measurements covering a wider range in $\lambda L_{\lambda}$.
Of course, adding these data points will be a challenging task given the complications of \civ-based reverberation mapping campaigns described in Section \ref{sect:intro}.

\subsection{Black hole mass}
\label{sect:mbh}
To determine the black hole mass, we need to know the width of the broad \civ\ emission line and the scale factor $f$ (Equation \ref{eqn:rm}).
The emission line width is measured on either the mean or root-mean-square (rms) spectrum, as the full width at half maximum (FWHM) or line dispersion, defined by
\begin{align}
\sigma_{\rm line}^2 = \left(\frac{\sum_i \lambda_i^2 F_i^2}{\sum_i F_i}\right) - \lambda_0^2,
\end{align}
where
\begin{align}
\lambda_0 = \frac{\sum_i \lambda_i F_i}{\sum_i F_i}.
\end{align}
Due to the strong absorption features on the core of \civ, estimating the peak in order to measure the FWHM is infeasible.
Similarly, while the line dispersion does not depend on an estimate of the emission line peak, the absorption will significantly affect this measurement.
Since neither measurement is possible on the data, we measure the emission line widths using the Gauss-Hermite fits to the broad \civ\ emission.

\begin{deluxetable}{ccccc}
\tablecaption{Emission line widths and black hole mass measurements}
\tablewidth{0pt}
\tablehead{ 
\colhead{} &
\multicolumn{2}{c}{Mean Spectrum} &
\multicolumn{2}{c}{rms Spectrum} \\
\colhead{Image} & 
\colhead{FWHM} &
\colhead{$\sigma_{\rm line}$} &
\colhead{FWHM} &
\colhead{$\sigma_{\rm line}$} \\ 
\colhead{} & 
\colhead{${\rm km~s}^{-1}$} &
\colhead{${\rm km~s}^{-1}$} &
\colhead{${\rm km~s}^{-1}$} &
\colhead{${\rm km~s}^{-1}$}
}
\startdata
A & $7925 \pm 100$ & $4499 \pm 80$ & $7705 \pm 372$ & $6100 \pm 286$ \\
B & $7555 \pm 93$ & $4209 \pm 69$ & $7848 \pm 665$ & $5477 \pm 199$ \\
C & $7700 \pm 96$ & $4018 \pm 87$ & $11400 \pm 860$ & $5774 \pm 202$ \\
combined & $7734 \pm 59$ & $4261 \pm 49$ & $9219 \pm 458$ & $5907 \pm 148$ \\
\hline\hline
 & \multicolumn{4}{c}{$\log_{10}(M_{\rm vir}/M_\odot) = \log_{10}(M_{\rm BH}/M_\odot) - \log_{10}f$} \\
\hline
A & $8.65 \pm 0.04$ & $8.16 \pm 0.04$ & $8.62 \pm 0.06$ & $8.42 \pm 0.06$ \\
B & $8.61 \pm 0.04$ & $8.10 \pm 0.04$ & $8.65 \pm 0.09$ & $8.33 \pm 0.05$ \\
C & $8.62 \pm 0.04$ & $8.06 \pm 0.05$ & $8.97 \pm 0.08$ & $8.37 \pm 0.05$ \\
combined & $8.63 \pm 0.04$ & $8.11 \pm 0.04$ & $8.79 \pm 0.06$ & $8.40 \pm 0.05$ \\
\hline\hline
 & \multicolumn{4}{c}{$\log_{10}(M_{\rm BH}/M_\odot)$} \\
\hline
A & $9.37 \pm 0.51$ & $8.67 \pm 0.26$ & --- & $9.05 \pm 0.14$ \\
B & $9.30 \pm 0.51$ & $8.63 \pm 0.26$ & --- & $8.96 \pm 0.13$ \\
C & $9.32 \pm 0.50$ & $8.57 \pm 0.27$ & --- & $9.00 \pm 0.13$ \\
combined & $9.35 \pm 0.51$ & $8.63 \pm 0.27$ & --- & $9.02 \pm 0.13$
\enddata
\tablecomments{ Emission line widths, virial product, and black hole mass measurements.
We give the measurements for each of the individual quasar images as well as the combined spectra.
The $f$ values used to calculate $M_{\rm BH}$ are $\log_{10} f_{{\rm CIV};{\rm rms},{\sigma}} = 0.63 \pm 0.12$, $\log_{10} f_{{\rm CIV};{\rm mean},{\sigma}} = 0.52 \pm 0.26$, and $\log_{10} f_{{\rm CIV};{\rm mean},{\rm FWHM}} = 0.69 \pm 0.50$.
We do not compute the value based on the FWHM and rms spectrum for the reasons described in Section \ref{sect:mbh}.
\label{table:linewidths}}
\end{deluxetable} 

To measure the emission line widths and estimate the uncertainties, we follow the Monte Carlo technique described by \citet{peterson04}, modified to fit our data set and spectral decompositions.
Given the $N$ epochs in the data set, we randomly select $N$ of these spectra, with replacement.
From each, we randomly select one \civ\ Gauss-Hermite fit from the posterior sample of the spectral decomposition for that night.
We then use this time-series of spectra to compute the mean and rms spectra and calculate the two line width measures.
We repeat this procedure for 1000 realizations and compute the mean and standard deviation of the distribution of measured line widths.
We perform this procedure for each quasar image individually and with the combined data set, the results of which are reported in Table \ref{table:linewidths}.

We pair the line width measurements with $\tau_{\rm cen} = 36.5^{+2.9}_{-3.9}$ from Section \ref{sect:lags} to determine the virial product, $M_{\rm vir} = M_{\rm BH}/f$, reported in Table \ref{table:linewidths}.
The correct conversion $f$ between $M_{\rm vir}$ and $M_{\rm BH}$ depends on a number of factors including the BLR geometry and kinematics, and will vary from one AGN to another.
Since this information is generally not available, one typically chooses $f$ such that the sample of reverberation mapped AGNs is aligned with quiescent galaxies in the $M-\sigma_*$ plane.
While numerous studies have calibrated $f$ to the $M-\sigma_*$ relation using the \Hb\ BLR, we are not aware of any that have done so for the \civ\ BLR.
Those that do calibrate \civ-based $M_{\rm BH}$ measurements do so using single-epoch methods with the $r_{\rm BLR}-L$ relation \citep[see, e.g.,][]{Park++17}, but the virial factor $f$ cannot be extracted from the fitted relations.

Since the geometry and kinematics of the two BLRs are not necessarily the same, and since \Hb\ and \civ\ emission line profiles tend to differ, the \Hb-based $f$ values may not be the same as those needed for \civ-based measurements.
However, since $M_{\rm BH}$ for an AGN is fixed regardless of which emission line is used to calculate it, we can use the relationship between the virial products $M_{{\rm vir},{\rm H}\beta}$ and $M_{{\rm vir,CIV}}$ to convert between $\log_{10}f_{{\rm H}\beta}$ and $\log_{10}f_{\rm CIV}$, according to
\begin{multline}
\log_{10}f_{\rm CIV} + \log_{10}(M_{{\rm vir,CIV}}/M_\odot) = \\\log_{10}f_{{\rm H}\beta} + \log_{10}(M_{{\rm vir},{\rm H}\beta}/M_\odot).
\end{multline}

\citet{Dallabonta++20} examined a sample of AGNs with $M_{\rm vir}$ available from both the \civ\ and \Hb\ emission lines and found that the two are consistent with each other when the line dispersion and mean spectrum are used.
Given that we have multiple line width measurements, we use the data they present to compute the offset $\alpha = \log_{10}(M_{{\rm vir},{\rm H}\beta}/M_\odot) - \log_{10}(M_{{\rm vir},{\rm CIV}}/M_\odot)$ for each pair of line with and spectrum so that $\log_{10}f_{\rm CIV} = \log_{10}f_{{\rm H}\beta} + \alpha$.
We find $\alpha_{{\rm mean},{\sigma}} = 0.087 \pm 0.007$, $\alpha_{{\rm rms},{\sigma}} = -0.021 \pm 0.028$, and $\alpha_{{\rm mean},{\rm FWHM}} = 0.694 \pm 0.008$, skipping $\alpha_{{\rm rms},{\rm FWHM}}$ since \citet{Dallabonta++20} do not provide those measurements.

For the \Hb-based $f_{{\rm rms},i}$ measurements, we use the values reported by \citet{Woo++15}, found by aligning a sample of reverberation mapped AGNs with the $M-\sigma_*$ relation for quiescent galaxies: $\log_{10} f_{{\rm H}\beta;{\rm rms},\sigma} = 0.65 \pm 0.12$ and $\log_{10} f_{{\rm H}\beta;{\rm rms}, {\rm FWHM}} = 0.05 \pm 0.12$.
For $f_{{\rm mean},i}$, we use the results of \citet{Williams++18} who examined a sample of 17 AGNs with BLR modeling: $\log_{10} f_{{\rm H}\beta;{\rm mean},\sigma} = 0.43 \pm 0.26$ and $\log_{10} f_{{\rm H}\beta;{\rm mean},{\rm FWHM}} = 0.00 \pm 0.50$.
Applying the offsets, we have $\log_{10} f_{{\rm CIV};{\rm rms},{\sigma}} = 0.63 \pm 0.12$, $\log_{10} f_{{\rm CIV};{\rm mean},{\sigma}} = 0.52 \pm 0.26$, and $\log_{10} f_{{\rm CIV};{\rm mean},{\rm FWHM}} = 0.69 \pm 0.50$.
We use these to compute the black hole masses reported in Table \ref{table:linewidths}.

We caution that the masses reported here are highly dependent on $f$ and the conversion between $f_{{\rm H}\beta}$ and $f_{\rm CIV}$.
Further research is necessary to determine the appropriate $f$ for more reliable calibration of \civ-based $M_{\rm BH}$ measurements.
The correct value can be determined through alignment of \civ\ reverberation mapped AGNs with the $M-\sigma_*$ relation or via direct modeling of the BLR, the latter of which will be the topic of a future paper with these data.

\begin{figure*}[t!]
\begin{center}
\includegraphics[width=7.0in]{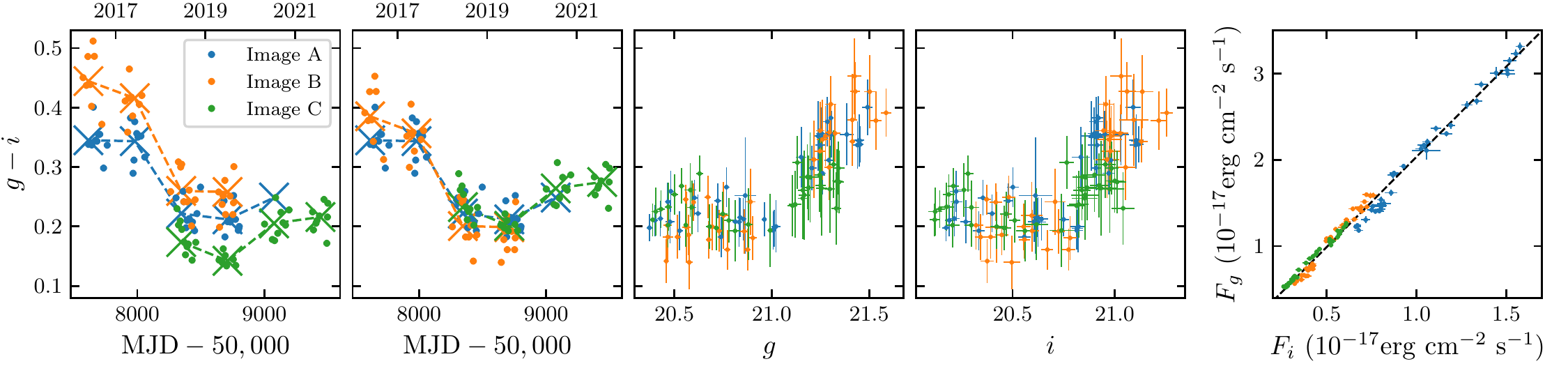}  
\caption{
\textit{Left}: Change in observed color over the duration of the observing campaign. The crosses mark the per-season mean $g-i$ for each image.
\textit{Middle-left}: Same as the left panel, but the three light curves have been corrected for differential extinction.
\textit{Middle and middle-right}: Change in observed color vs. g- and i-band photometry. The $i$-band has been corrected for differential extinction.
\textit{Right}: Comparison of $g$- and $i$- band fluxes. A linear fit to the data is shown by the dashed black line.
\label{fig:gminusi}}
\end{center}
\end{figure*}

\subsection{Correlated variability of observed color and luminosity}
Since the light for the three quasar images passes through different parts of the lensing galaxy cluster, each image experiences a different amount of extinction.
We can assess this by comparing the $g-i$ photometry for the three images on matching dates.
First, we take the dates in which we obtained photometry and shift them by the measured lensing time delays.
We then group the $g$- and $i$-band photometry by observing season and compute the median $g-i$ for each image.
For the seasons in which data overlap, we compute the $g-i$ offsets between the images and then calculate the mean offset relative to image A to find $(g-i)_A - (g-i)_B = -0.059$ and $(g-i)_A - (g-i)_C = 0.059$.
The $g-i$ light curves are shown in the left panel of Figure \ref{fig:gminusi}, where the crosses indicate the median $g-i$ for each image and observing season.
In the middle-left panel, we shift images B and C by the measured offsets to align with image A.

In the middle and middle-right panels, we show $g-i$ plotted against the $g$- and $i$-band photometry.
Here, the $g$- and $i$-band photometry for images B and C have been shifted according to the measured relative magnifications $\Delta m_{A,B}$ and $\Delta m_{A,C}$, and the $i$ band has been adjusted according to the above measured $g-i$ offsets.
There is a clear trend in which $g-i$ increases with both $g$ and $i$ for $g>21.0$ and $i>20.5$.
In the right-hand panel we show the $g$-band flux plotted against the $i$-band flux.
The data are well fit by straight line $F_g = (2.09 \pm 0.03) (F_i - 10^{-17}~{\rm erg~s}^{-1}~{\rm cm}^{-2}) + (2.03 \pm 0.02)\times 10^{-17}~{\rm erg~s}^{-1}~{\rm cm}^{-2}$.
This indicates that the change in $g-i$ is not due to a change in the AGN power law continuum slope, but rather the effect of AGN variability over a constant contribution from the red host galaxy.

\subsection{Systematic offset in the image B \civ\ emission line}
\label{sect:microlensing}
The image magnifications reported by \citet{Dahle++15} are measured using the $g$-band continuum light curve.
However, when plotting the emission line light curves using these magnifications, we find that the image B light curves lie systematically below the two other light curves.
We attribute this effect to differential microlensing which will affect the small-scale accretion disk, but not the more extended broad line region.

To quantify the effect, we use the narrow-band \civ\ light curves from spectroscopy.
First, we linearly interpolate the image B light curve onto the image A light curve time values.
We simply linearly interpolate the error bars, so we account for additional uncertainty in the interpolation by introducing a parameter $\delta$ such that 
\begin{align}
\sigma_{B^\prime,{\rm tot}}^2 = \sigma_{B^\prime}^2 + (\delta B^\prime)^2,
\end{align}
where $B^\prime$ is the interpolated light curve, $\sigma_{B^\prime}$ is the interpolated uncertainty, and $\sigma_{B^\prime,{\rm tot}}^2$ is the total uncertainty on $B^\prime$.
We then maximize the log-likelihood
\begin{align}
l(\alpha, \delta) = -\frac{1}{2} \sum_{i}\left[\frac{(A_i - \alpha B_i)^2}{\sigma_{{\rm tot},i}^2(\delta)} - \ln (2 \pi \sigma_{{\rm tot},i}^2[\delta])\right]
\end{align}
where $\sigma_{{\rm tot},i}^2(\delta) = \sigma_{A,i}^2 + \sigma_{B^\prime,{\rm tot},i}^2(\delta)$, and $\alpha$ is the scale factor required to bring $B$ into alignment with $A$.
Using the code {\sc emcee} \citep{Foreman-Mackey++13}, we measure the median and 68\% confidence intervals of the MCMC chains for the parameter $\alpha$ to find $\alpha = 1.179^{+0.010}_{-0.011}$, with $\log_{10}\delta=-3.0\pm0.2$.
We repeat this same process again, instead interpolating the image A light curve onto the image B light curve, and find $\alpha = 1.181^{+0.011}_{-0.012}$, with $\log_{10}\delta=-2.8\pm0.2$.
We then perform the same procedure using the continuum-subtracted narrow-band photometry light curves to get $\alpha = 1.066^{+0.016}_{-0.016}$ and $\alpha = 1.069^{+0.017}_{-0.018}$, each with negligible $\delta$.

These correction factors should match each other since the two sets of light curves are measuring the same thing.
There are two potential sources for the discrepancy: (1) if the photometry aperture and slit allow different amounts of AGN or host-galaxy light through to the detector, or (2) if the two continuum subtraction methods have a systematic offset.
We do not expect (1) to be the issue since we use a large ($2^{\prime\prime}$) slit to avoid any slit losses.
Examining the raw, continuum-unsubtracted spectra and photometry confirms that both narrow-band measurements are closely aligned.
It is more likely that (2) is the issue since the continuum estimate from photometry is based solely on the $g$- and $i$-band photometry.
While the continuum estimates under \civ\ appear to be in good agreement with the $g$-band continuum light curve, we don't take these measurements to be as reliable as the spectra-based measurements due to the larger uncertainties involved.

We apply an $\alpha=1.180$ correction to all emission line light curves for image B before computing the cross-correlation functions.
We also perform tests in which we use $\alpha=1.068$ for the narrow-band photometry, but find that the lag measurements are consistent to within the uncertainties.

\section{Summary}
We have presented the first results of a 4.5 year spectroscopic and photometric monitoring campaign of the multiply imaged quasar SDSS J2222+2745 at $z=2.805$.
Our main results can be summarized as follows:
\begin{enumerate}
\item After performing percent-level spectrophotometric flux calibration, we produce emission line light curves spanning over six years in duration in the observed frame. We give light curves for the integrated \civ\ emission line, a narrow band filter covering the \civ\ core, and four wavelength ranges covering the blue and red wings of \civ.

\item We measure integrated and velocity-resolved \civ\ emission line lags with respect to the $g$-band continuum. The integrated \civ\ lag is $\tau_{\rm cen} = 36.5^{+2.9}_{-3.9}$ days in the rest frame ($139^{+11}_{-15}$ days observed frame). We see velocity-resolved lag structure in which the core of the line responds the slowest ($86.2^{+4.5}_{-5.0}$ days) and the wings respond the fastest ($25^{+11}_{-15}$ and $7.5^{+4.2}_{-3.5}$ days for the blue and red wings, respectively). This behavior is consistent with BLR gas that is in circular Keplerian motion.

\item We place SDSS J2222+2745 on the $r_{\rm BLR}-L$ relation with 33 data points from the literature and fit a linear regression to the data, finding $\log_{10}(\tau / {\rm days}) = (1.00 \pm 0.08) + (0.48 \pm 0.04) \log_{10}[\lambda L_\lambda(1350{\rm \AA}) / 10^{44}~{\rm erg~s}^{-1}]$ with an intrinsic scatter of $0.32 \pm 0.06$ dex. SDSS J2222+2745 lies 0.58 dex above the mean relation.

\item We calculate emission line widths and pair these with the measured lags to obtain the virial product, $M_{\rm vir} = M_{\rm BH}/f$, for the four combinations of the (mean, rms) spectrum and (FWHM, $\sigma_{\rm line}$) emission line widths. Assuming values of $f$ based on \Hb, we compute the black hole mass for each combination.

\end{enumerate}

Future analyses with these data will include measurements of the broad \ciii\ $\lambda$1909 emission line.
We aim to extend this campaign through the end of 2022 for a full campaign duration of 6.5 years (8.5 years with gravitational time delays), which will provide the data for more detailed analyses of the geometry and kinematics of the \civ\ and \ciii\ BLRs, and an absolute calibration of the $f$ factor and the inferred black hole mass.


\acknowledgments

The data presented here are based in part on observations obtained at the international Gemini Observatory, a program of NSF’s NOIRLab, which is managed by the Association of Universities for Research in Astronomy (AURA) under a cooperative agreement with the National Science Foundation on behalf of the Gemini Observatory partnership: the National Science Foundation (United States), National Research Council (Canada), Agencia Nacional de Investigaci\'{o}n y Desarrollo (Chile), Ministerio de Ciencia, Tecnolog\'{i}a e Innovaci\'{o}n (Argentina), Minist\'{e}rio da Ci\^{e}ncia, Tecnologia, Inova\c{c}\~{o}es e Comunica\c{c}\~{o}es (Brazil), and Korea Astronomy and Space Science Institute (Republic of Korea).
The Gemini data were obtained from programs GN-2016B-Q-28, GN-2017A-FT-9, GN-2017B-Q-33, GN-2018A-Q-103, GN-2018B-Q-143, GN-2019A-Q-203, GN-2019B-Q-232, GN-2020A-Q-105, and GN-2020B-Q-132 (PI Treu), and were processed using the Gemini IRAF package.

The data presented here were obtained in part with ALFOSC, which is provided by the Instituto de Astrofisica de Andalucia (IAA) under a joint agreement with the University of Copenhagen and NOTSA.

This work was enabled by observations made from the Gemini North telescope, located within the Maunakea Science Reserve and adjacent to the summit of Maunakea. We are grateful for the privilege of observing the Universe from a place that is unique in both its astronomical quality and its cultural significance.

This research made use of Astropy,\footnote{http://www.astropy.org} a community-developed core Python package for Astronomy \citep{astropy:2013, astropy:2018}

PW and TT gratefully acknowledge support by the National Science Foundation through grant AST-1907208 ``Collaborative Research: Establishing the foundations of black hole mass measurements of AGN across cosmic time''  and by the Packard Foundation through a Packard Research Fellowship to TT.
Research at UCI Irvine was supported by NSF grant AST-1907290. 
KH acknowledges support from STFC grant ST/R000824/1.

\bibliographystyle{apj}
\bibliography{references}

\end{document}